\documentclass[journal,draftclsnofoot,onecolumn,12pt]{IEEEtran}%

\usepackage{amsmath}
\usepackage{tikz}

\usepackage{amssymb}

\usepackage{tabu}
\usepackage{todonotes}
\usepackage{epsf}
\usepackage{epsfig}
\usepackage{times}
\usepackage{epsfig}
\usepackage{graphicx}
\usepackage{bbold}
\usepackage{mathtools}
\usepackage{mathrsfs}
\usepackage{amssymb,amsmath}
\usepackage{pdfpages}

\usepackage[outdir=./]{epstopdf}

\usepackage{dsfont}
\usepackage{lettrine} 

\usepackage{amsmath,epsfig,amssymb,amsthm,cite,url}

\usepackage[justification=centering]{caption}
\usepackage{cite}
\usepackage{bbm}
\usepackage{algorithmic}
\usepackage{algorithm}

\allowdisplaybreaks
\usepackage{csquotes}
\usepackage{amsmath,amssymb}

\usepackage[english]{babel}
\usepackage{amsmath,amssymb}

\newtheorem{corollary}{Corollary}
\let\oldcorollary\corollary
\renewcommand{\corollary}{\oldcorollary\normalfont}
\newtheorem{theorem}{\bf Theorem}
\let\oldtheorem\theorem
\renewcommand{\theorem}{\oldtheorem\normalfont}

\newtheorem{proposition}{\bf Proposition}
\let\oldproposition\proposition
\renewcommand{\proposition}{\oldproposition\normalfont}
\newtheorem{lemma}{\bf Lemma}
\let\oldlemma\lemma
\renewcommand{\lemma}{\oldlemma\normalfont}
\newtheorem{example}{Example}
\let\oldexample\example
\renewcommand{\example}{\oldexample\normalfont}

\newtheorem{definition}{\bf Definition}
\let\olddefinition\definition
\renewcommand{\definition}{\olddefinition\normalfont}
\newtheorem{remark}{Remark}
\let\oldremark\remark
\renewcommand{\remark}{\oldremark\normalfont}

\usepackage{changes}
\usepackage{acronym}

\usepackage{color}
\usepackage{xcolor}
\usepackage{dsfont}

\newcommand{\rhob}{\boldsymbol{\rho}}
\newcommand{\Pb}{\boldsymbol{P}}

\newcommand{\lambdab}{\boldsymbol{\lambda}}

\usepackage{tikz}
\usetikzlibrary{arrows}

\newcommand{\argmin}{\text{arg}\min}
\newcommand{\Ex}{\mathds{E}}

\newcommand{\psb}{{\boldsymbol{\psi}}}

\usepackage{tikz}
\usetikzlibrary{arrows}

\newcommand{\rb}{\boldsymbol{r}}
\newcommand{\Pro}{\text{Pr}}
\newcommand{\ab}{\boldsymbol{a}}
\newcommand{\stb}{\boldsymbol{s}}

\newcommand{\tr}{\textrm{tr}}
\newcommand{\VAR}{\text{VAR}}
\newcommand{\COV}{\text{COV}}
\newcommand{\mysize}{0.50}

\DeclareMathOperator{\psim}{g_{\text{sim}}}

\usepackage{setspace}	

\usepackage{makecell}
\usepackage[font=small,labelfont=bf]{caption}

\begin{document}
\newacro{drl}[deep-RL]{deep reinforcement learning}
\newacro{qoe}[QoE]{quality-of-experience}
\newacro{qos}[QoS]{quality-of-service}
\newacro{gan}[GAN]{generative adversarial network}
\newacro{rb}[RB]{resource block}
\newacro{mdp}[MDP]{Markov decision process}
\newacro{ofdma}[OFDMA]{orthogonal frequency-division multiple access}
\newacro{dnn}[DNN]{deep neural network}
\newacro{bs}[BS]{base station}
\newacro{ppo}[PPO]{proximal policy optimization}
\newacro{dqn}[DQN]{deep Q-networks}
\newacro{sgd}[SGD]{stochastic gradient descent}
\newacro{urllc}[URLLC-6G]{ultra reliable low latency communication}

	\newcommand{\bb}[1]{\mathbb{#1}}
	\newcommand{\mc}[1]{\mathcal{#1}}
	%
	%
	\title{\LARGE Experienced Deep Reinforcement Learning with Generative Adversarial Networks (GANs) for Model-Free Ultra Reliable Low Latency Communication}

\author{\IEEEauthorblockN{  Ali Taleb Zadeh Kasgari~\IEEEmembership{Student Member,~IEEE}, Walid Saad,~\IEEEmembership{ Fellow,~IEEE}, Mohammad Mozaffari~\IEEEmembership{Member,~IEEE}, and H. Vincent Poor,~\IEEEmembership{ Fellow,~IEEE}}\\
	\IEEEauthorblockA{
	\thanks{This  work  was  supported by  the  U.S. National Science Foundation under Grants IIS-1633363 and CNS-1836802. The work of H. V. Poor was supported by  the  U.S. National Science Foundation under Grant CCF-1908308}
		\thanks{A. Taleb Zadeh Kasgari and W.~Saad are with Wireless@VT, Department of ECE, Virgina Tech, Blacksburg, VA, 24060, USA. Emails: \{alitk, walids\}@vt.edu. M. Mozaffari is with Ericsson Research, Santa Clara, CA, 95054, USA, Email: {mohammad.mozaffari@ericsson.com}. H. V. Poor is with the Department of Electrical Engineering, Princeton University, Princeton, NJ, 08544, USA, Email: {poor@princeton.edu}.}
        \thanks{A preliminary version of this work appeared in IEEE ICC, \cite{Tale1905:Model}.}
	}}

	\maketitle
	
\vspace{-4em}
	\begin{abstract}
In this paper, a novel \emph{experienced} deep reinforcement learning (deep-RL) framework is proposed to provide model-free resource allocation for \ac{urllc} in the downlink of a wireless network. The goal is to guarantee high end-to-end reliability and low end-to-end latency, under explicit data rate constraints,  for each wireless user without any models of or assumptions on the users' traffic. In particular, in order to enable the deep-RL framework to account for extreme network conditions and operate in highly reliable systems, a new approach based on generative adversarial networks (GANs) is proposed. This GAN approach is used to pre-train the deep-RL framework using a mix of real and synthetic data, thus creating an experienced deep-RL framework that has been exposed to a broad range of network conditions. The proposed deep-RL framework is particularly applied to a multi-user  orthogonal frequency division multiple access (OFDMA) resource allocation system.
  Formally, this \ac{urllc} resource allocation problem in OFDMA systems is posed as a power minimization problem under reliability, latency, and rate constraints. To solve this problem using experienced deep-RL, first, the rate of each user is determined. Then, these rates are mapped to the resource block and power allocation vectors of the studied wireless system. Finally, the end-to-end reliability and latency of each user are used as feedback to the deep-RL framework. It is then shown that at the fixed-point of the deep-RL algorithm, the reliability and latency of the users are near-optimal. 
 Moreover, for the proposed GAN approach, a theoretical limit for the generator output is analytically derived. 
 Simulation results show how the proposed approach can achieve near-optimal performance within the rate-reliability-latency region, depending on the network and service requirements. The results also show that the proposed experienced deep-RL framework is able to remove the transient training time that makes conventional deep-RL methods unsuitable for \ac{urllc}. Moreover, during extreme conditions, it is shown that the proposed, experienced deep-RL agent can recover instantly while a conventional deep-RL agent takes several epochs to adapt to new extreme conditions. 
	\end{abstract}


	%
	\IEEEpeerreviewmaketitle
	\vspace{-1.5em}
	\section{Introduction}
	Ultra reliable low latency communication will be one of the most important features in  next-generation 5G and 6G cellular networks as it will be necessary for mission critical applications such as Internet of Things (IoT)~\cite{Mozaffari2017IoT} sensing and control as well as remote control of autonomous vehicles and drones \cite{bennis2018ultra, saad2019vision}. Thus far, prior research in this area has been mostly focused on  applications that require low data rates such as uplink and short packets transmissions of IoT sensors\cite{bennis2018ultra, rahmati2018survey}. 
    However, new wireless applications such as drone communications \cite{Mozaffari2017IoT}, autonomous driving \cite{ye2018deep,ferdowsi2018robust}, and virtual reality \cite{chen2018virtual}, have emerged. These applications require reliable and low latency communications, not only in the uplink, but also in the downlink for control, navigation, and tracking purposes. {Moreover, in order to operate effectively, such applications require high reliability, low latency, and certain data rate guarantees. For example, autonomous vehicles will need to receive reliable but high data rate information such as HD maps from infrastructure. Similarly, virtual and extended reality applications \cite{chaccour2020can} will require high data rates with high reliability for transmitting, in real-time, the virtual environment videos and images. These applications will need a mix of low latency and reliability guarantees as well as enhanced mobile broadband (eMBB) data rates. Moreover, in these applications the data packets will not be simple short packet transmissions as is the case in classical 3GPP ultra reliable low latency communications (URLLC). Instead, these applications will generate large packets due to the nature of their content \cite{ren_joint_2020,jurdi_variable-rate_2018}.}  {Therefore, we define a new type of service, that we call \emph{URLLC-6G}, which is different from classical 3GPP URLLC, in that it is suitable to a broader range of applications that must provide highly reliable and extremely low latency communication links to wireless users without imposing any specific limitations on  their packet size and rate requirements.}

    {Providing highly reliable and low latency communications with rate considerations for these \ac{urllc} applications poses many network challenges.}
    First, such applications are very sensitive to wireless network environment fluctuations and, therefore, they require wireless resource management solutions that are reliable in face of \emph{extreme network conditions} (e.g., unexpected traffic patterns or worst-case randomness of the wireless channel). Second, considering the limited radio resources in a communication system, low latency, high reliability, and high rate can become three incompatible design parameters. This incompatibility means that improving one of them could potentially be detrimental to the other two, thus requiring \emph{new designs that can be used to balance the rate-reliability-latency tradeoff} for the aforementioned applications~\cite{bennis2018ultra}. Third, maintaining high reliability and low latency needs timely and efficient resource allocation. Hence, \ac{urllc} resource allocation should allocate the exact amount of resources required by the users. In other words, in order to provide \ac{urllc} links, the resource management system of a cellular system must be able to sustain extreme network conditions. Moreover, to balance the rate-reliability-latency tradeoff, any resource allocation scheme must learn each user's exact performance requirements so that it can satisfy them without wasting any resources or reducing the user's reliability.

\subsection{Related Works}
    Recently, there has been a surge in literature that studies problems of URLLC and resource allocation, such as in \cite{Bennis2018EVT,Hou2018Bursty, lai2013qos, Zhou2018VTC, bennis2018ultra,yousefvand2018joint,samarakoon2018federated,ashraf2018access,  Kasgari2019TCOM, anand2018infocom}.
    In \cite{Bennis2018EVT}, the authors
    use extreme value theory to study URLLC in a vehicular network and characterize the queue statistics.      
    The work in \cite{Hou2018Bursty} considers a model-based and a data-driven approach for designing a burstiness-aware scheduling framework which reserves bandwidth for users with bursty traffic. 
    The authors in \cite{lai2013qos} propose a packet prediction mechanism to predict the behavior of future incoming packets based on the packets in the current queue.
    In \cite{Zhou2018VTC}, the authors propose a hybrid resource allocation, to allocate dedicated resources and a shared resource pool to a set of wireless users with 3GPP URLLC requirements considering channel and users' traffic conditions. Meanwhile, the work in \cite{yousefvand2018joint} introduces a method for joint user association and resource allocation with reliability considerations. A joint transmit power and resource allocation for vehicular networks  subject
to ultra reliability and low latency constraints is proposed in \cite{samarakoon2018federated}. The work in \cite{ashraf2018access} proposes a quality-of-service-aware resource allocation framework for vehicular networks. In \cite{anand2018infocom}, an algorithm for joint scheduling of URLLC and broadband traffic in 5G cellular systems is proposed. In \cite{Kasgari2019TCOM}, we studied the problem of providing low latency communication for human-centric applications considering the reliability of human users.
    However, all of these existing works assume that explicit traffic and queue models are available to the resource allocation system \cite{Bennis2018EVT,Hou2018Bursty, lai2013qos, Zhou2018VTC, bennis2018ultra,yousefvand2018joint,samarakoon2018federated,ashraf2018access,  Kasgari2019TCOM, anand2018infocom}. Since the assumed models are often simplified, they either underestimate or overestimate the users' traffic and queue lengths. This can cause the resource allocation algorithm to either allocate more resources or less resources than the actual requirement of the users which, in turn, can render the system inefficient or degrade the reliability of the users' connections. Moreover, prior works which consider extreme cases for URLLC \cite{Bennis2018EVT} with a simplified traffic model, cannot handle practical extreme network scenarios which further motivates a model-free approach. Further, the previous works on URLLC completely ignore any rate requirements of the users and some of them such as \cite{Hou2018Bursty} and \cite{lai2013qos} heavily rely on historical data which can also lead to inefficient resource allocation because a user's traffic often changes based on spatial or temporal factors. In fact, even if it is available, historical data is often not a precise predictor of the traffic of wireless users.

     
    To overcome some of these challenges, there has been recent interest in using deep reinforcement learning (deep-RL) for solving wireless networking problems with incomplete information  \cite{He:2017:DRL,Xu2017ICC,Li2018WCNC,ye2018deep,ahmed2018deep,sun2017spawc,azari2018ML}. In \cite{ye2018deep}, a decentralized resource allocation framework for vehicle-to-vehicle communications is proposed based on deep-RL. The authors in \cite{He:2017:DRL} propose a deep-RL resource management approach for virtualized ad-hoc network. In \cite{minghze2017JSACcaching} a deep-RL resource management system is proposed for caching in cloud radio access networks. A deep-RL based resource allocation scheme for mobile edge computing is studied in \cite{Li2018WCNC}. The work in \cite{ahmed2018deep} proposes a deep supervised learning approach to solve the sub-band and power allocation
problem in multi-cell networks. A deep-RL based distributed pilot sequence selection algorithm is proposed in \cite{Wong2019Globecom}. In \cite{sun2017spawc}, the authors propose a learning-based approach for wireless resource management by modeling the input-output relationship of a resource optimization algorithm with neural network.
    However, these works do not investigate \ac{urllc} problems as they focus on 3GPP URLLC. Moreover, since these deep-RL works \cite{Xu2017ICC,ahmed2018deep,Li2018WCNC} limit their problem's action space, use discretization to manage the size of the action space \cite{ye2018deep}, or do not address the limitation of deep-RL when dealing with large action spaces, they are not suitable for realistic \ac{urllc} resource management problems in actual wireless systems.  This is because these works cannot handle the large action space involved in \ac{urllc} and they have high order of time complexity which makes them not suitable for the requirements of \ac{urllc}.

    Collecting real data for training deep learning models is another challenge which is not properly addressed in the previous deep-RL  works that mostly rely on simulated data \cite{He:2017:DRL,Xu2017ICC,Li2018WCNC,ye2018deep,ahmed2018deep,azari2018ML,sun2017spawc, Wong2019Globecom, He2017TVT}. 
    Although deep-RL methods do not need collecting data beforehand, a deep-RL agent can better perform if it is trained before being deployed. This is particularly needed for \ac{urllc} systems in which reliability is necessary. Indeed, even though
deep-RL methods can collect data during the training process \cite{deepRL2015}, as is true in the case of humans, a deep-RL agent can substantially enhance its prediction if it becomes more intelligent and experienced by exploiting more training datasets. Hence, if a deep-RL agent is deployed for \ac{urllc} purposes without training, it will initially be inexperienced at the beginning of the learning which can cause unreliability in the system. Most of the previous literature on deep-RL did not consider this issue of learning reliability  \cite{He:2017:DRL,Xu2017ICC,Li2018WCNC,ye2018deep,ahmed2018deep,azari2018ML} and hence are not easily applicable for future \ac{urllc} applications.   In a \ac{urllc} system, \emph{intelligence and experience} are particularly important to enable the deep-RL agent to deal with \emph{extreme and critical events} that occur in the wireless system and that can jeopardize the performance of the system. Examples of such extreme events include unusual traffic patterns, extremely scarce resource availability, unforeseen network congestion, and deep fades that seldom occur, among others. In fact, the whole premise of \ac{urllc} is to provide reliability with respect to such extreme events \cite{anand2018infocom,Bennis2018EVT,Hou2018Bursty,lai2013qos,bennis2018ultra,yousefvand2018joint,samarakoon2018federated,ashraf2018access}. However, without gaining considerable experience through rigorous training, it becomes unlikely for deep-RL agents to frequently encounter such extreme events and, thus, they cannot cope with \ac{urllc} requirements. As a result, prior works on deep-RL~ \cite{He:2017:DRL,Xu2017ICC,Li2018WCNC,ye2018deep,ahmed2018deep,sun2017spawc,azari2018ML, Wong2019Globecom} that solely rely  on online training may not be able to provide \ac{urllc} links at the level expected by 5G systems (i.e., very high reliability at very low latency). 
    In order to overcome this unique \ac{urllc} challenge, we propose to develop an \emph{``experienced'' deep-RL agent} that can gain experience in a virtual environment before being deployed in an extremely reliable systems.

\subsection{Contributions}
\begin{figure}[!t]
    \centering
    \includegraphics[width=.95\textwidth]{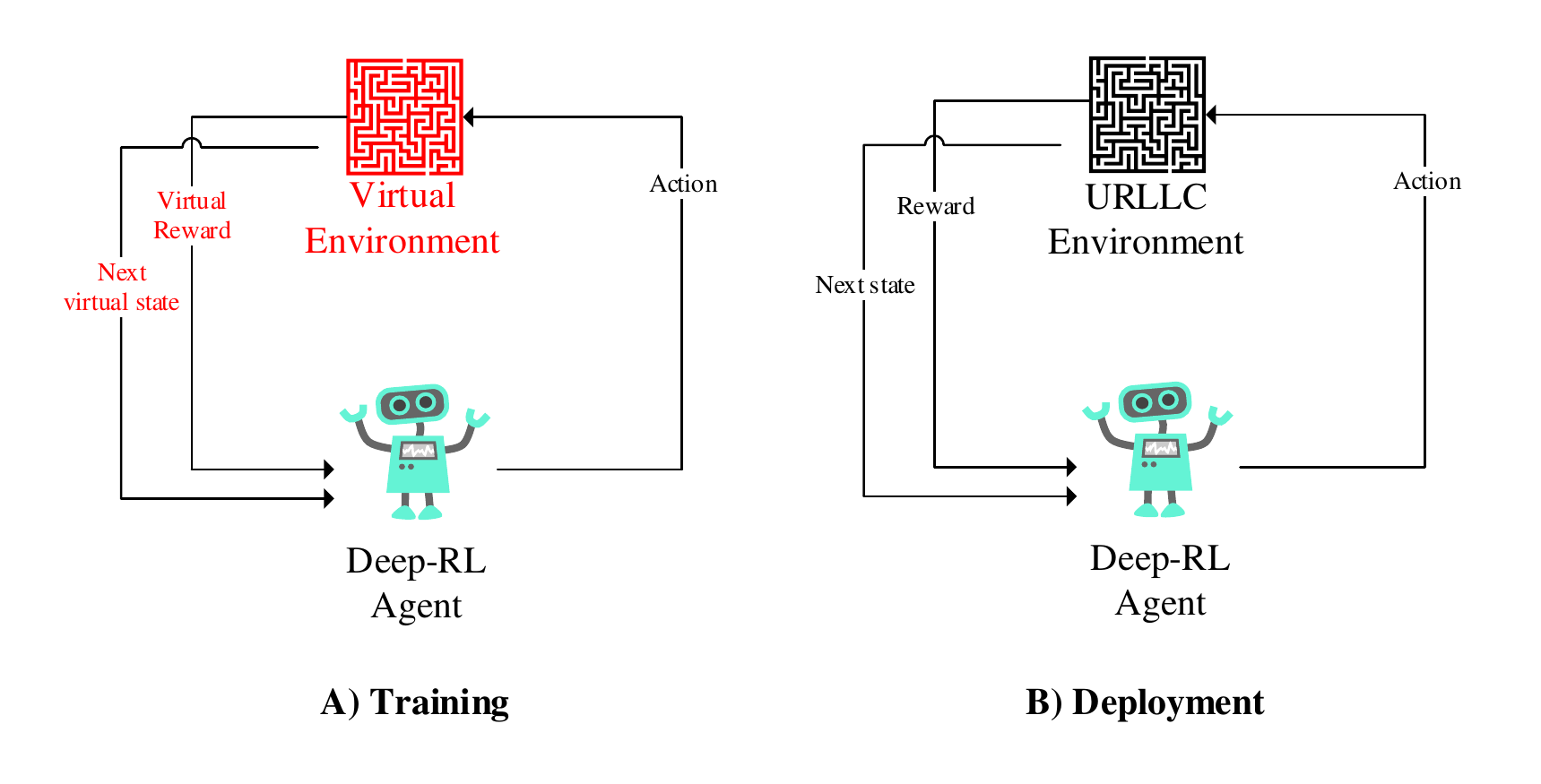}
    \vspace{-2.2em}
    \caption{Providing deep-RL with network experience in a virtual environment before deployment in a real \ac{urllc} environment.}
    \label{fig:VirtualEnvironment}
    \vspace{-2em}
\end{figure}

    The main contribution of this paper is  
     a novel, model-free resource management framework that is reliable against extreme events and that can balance the tradeoff between reliability, latency, and data rate, without explicit prior assumptions on the system parameters such as the users' traffic arrival model or the wireless environment.  We formulate the problem as a model-free power minimization problem with reliability, latency, and rate considerations. 
     To solve this problem, we propose a deep-RL framework that dynamically predicts the traffic model of the users and, then, uses those predictions to jointly allocate resource blocks (RBs) and power to downlink users, under \ac{urllc} and rate constraints. {Although \ac{drl} methods can normally collect data during the learning process, and they do not need a dataset of labeled data beforehand, this can cause unreliability in the \ac{drl} controllers. Therefore, using this classical approach for \ac{drl} is not suitable for ultra reliable low latency communication systems in which reliability is of critical importance.} Therefore, we introduce a \emph{generative adversarial network (GAN)-based refiner} to that can enable the proposed deep-RL agent to gain experience and learn extreme events. The proposed GAN-based refiner allows us to transform a limited set of real traffic and wireless channel data into a large dataset of realistic and diverse data that can be deployed for training the deep-RL agent and providing it with experience. Fig. \ref{fig:VirtualEnvironment} illustrates how we propose to use a virtual environment to increase the deep-RL agent's experience. Clearly, the deep-RL agent will act in a virtual training environment and using virtual feedbacks it gains experience. Then, it is deployed in an actual \ac{urllc} environment.  The proposed framework is then shown to effectively find a near-optimal resource allocation solution such that the low latency, high reliability, and high rate requirements of the wireless users are satisfied.  Also, it is shown that the GAN-generated dataset can enable the framework to gain experience in extreme events which allows it recover faster in the case of unpredicted  rare events in a \ac{urllc} system.   The term ``actual environment" is here used to differentiate between the virtual environment which is an offline training environment, and a real simulated environment where we are testing our \ac{drl} agent.

     In summary, our key contributions include:
    \begin{itemize}
    \item The proposed deep-RL framework can dynamically can measure the end-to-end reliability and the delay of each user. Then, it uses this measurement as online feedback to modify its decisions. In particular, the deep neural network (DNN) weights used in deep-RL are updated using this feedback only.
   Also, the proposed resource allocation system is then shown to be able to predict the consequences of its actions in the future and use this information to make better resource allocation decisions.  This helps the algorithm provide long-term reliability and latency guarantees for the users.  
   
   \item 
    Unlike the deep-RL approaches that were previously used for wireless networks, e.g., in \cite{He:2017:DRL,Xu2017ICC,Li2018WCNC,ye2018deep,ahmed2018deep}, and \cite{azari2018ML},  that use an inexperienced deep-RL agent our proposed deep-RL framework is designed so as to gain network experience in a virtual environment generated by using the proposed GAN-based refiner.  The use of GANs enables our deep-RL framework to avoid trial-and-errors imposed by other approaches, thus making our solution suitable for \ac{urllc} scenarios.  Creating this GAN-based virtual environment can also remove biases in the wireless datasets that stem from the fact that wireless channel information and packet arrival information are normally gathered in a specific time and location which makes them biased towards those spatio-temporal conditions.

    \item
   The proposed GAN model can use a real dataset with limited data points on the packet arrival and wireless channel information to generate a comprehensive real dataset of packet arrival and channel information that can include extreme networking conditions. Then, the GAN-generated dataset, combined with the real dataset, can be fed into the deep-RL agent, to create an \emph{experienced deep-RL agent} that can provide \ac{urllc}. To the best of our knowledge, none of the prior works in \cite{He:2017:DRL,Xu2017ICC,Li2018WCNC,ye2018deep,ahmed2018deep}, and \cite{azari2018ML,deepRL2015,lillicrap2015continuous,PPOmain,PPOcontinous} developed an experienced deep-RL agent that makes use of GAN to learn extreme events and eliminate data set biases. We note that some works in wireless literature \cite{ye2018GAN,ye2019GAN}, and \cite{AidinGlobecom} have used GAN. However, these works used GAN to solely identify the distribution of the wireless channel or for security purposes. Although our work identifies the channel and traffic distribution implicitly, it also gives the generator network the ability to generate extreme cases -- an aspect that has  not been investigated in any of previous works  \cite{ye2018GAN,ye2019GAN}. We also analytically find a range for the generator network's extreme cases.
   
        \item We enhance deep-RL scalability by addressing the large action space problem that stems from the fact that in the considered wireless network, there is a large set of actions which can be taken by deep-RL agent. This large set of actions makes the problem unsuitable for deep-RL frameworks \cite{deepRL2015}.  In particular, we propose the novel concept of an \emph{action space reducer} which reduces the size of the action space without limiting it.
     Using this action space reducer, our deep-RL framework can make decisions in real-time as opposed to discretization approach used in \cite{ye2018deep}. We show that when the proposed algorithm converges, the reliability, latency, and rate of each user are guaranteed.
    \end{itemize}

   Simulation results show that the proposed, experienced deep-RL framework can ensure reliability by eliminating the transient training time in the conventional deep-RL method. Moreover, during extreme conditions, our experienced deep-RL agent can recover almost instantly while it takes 50 epochs for a conventional deep-RL agent to adapt to new extreme conditions. This makes our proposed deep-RL agent more suitable for \ac{urllc} systems.

   The rest of the paper is organized as follows. Section
\ref{sec:SysModel} introduces the system model. Sections \ref{sec:deepRLURLLC} and \ref{sec:GAN}, respectively, present the deep-RL and GAN-based virtual environment for model free \ac{urllc}.
	Section \ref{sec:simulResults} presents the simulation results and conclusions are drawn
	in \ref{sec:conclusion}.

\section{System Model}\label{sec:SysModel}
Consider the downlink of an orthogonal frequency-division multiple access (OFDMA) cellular network with a single base station (BS)  serving a set $\mathcal{N}$ of $N$ users and having a set $\mathcal{K}$ of $K$ available RBs.  {Although, we consider a single cell base-station, our results can be further generalized by what we mention in Subsection \ref{sec:ActionReducer}}  Each user has its own, individual rate, reliability, and latency requirements. We do not make any assumptions on the packet arrival or the packet length of each user.
The downlink transmission rate from the BS to user $i$ is:
\begin{equation}
r_i(t)=\sum_{j=1}^K \rho_{ij}(t) B \log_{2}\left(1+ \frac{p_{ij}(t)h_{ij}(t)}{\sigma^2}\right),
\end{equation}
where $B$ is the RB bandwidth, and $h_{ij}(t)$ is  the time-varying  channel gain of the transmission from the BS to user $i$  on RB $j$ at time slot $t$. $p_{ij}(t)$ is the downlink transmission power of the BS on RB $j$ to user $i$ at  slot $t$.  $\rho_{ij}(t)$ is the RB allocation indicator with $\rho_{ij}(t)=1$ when RB $j$ is allocated to user $i$ at time slot $t$, otherwise $\rho_{ij}(t)=0$. $\sigma^2=B N_0$ is the noise power with $N_0$ being the power spectral density of noise. 
 {The rate in (1) can be modified to accomodate other models such as the finite blocklength regime \cite{poor_blocklength_2010} in order to support classical short packet \ac{urllc} scenarios such as those used for factory automation. However, this would not change our general system design, but it will just impact the action space reducer component that we discuss in Section \ref{sec:ActionReducer}}.
 The resource allocation system will decide on the composition of the users in each row (time slot).

We define \emph{reliability} $\gamma_i(t)$, according to to the official 3GPP definition in \cite{3gpp}, as the probability of the end-to-end instantaneous packet delay exceeding a predefined target end-to-end latency threshold $D_i^{\max}$ for user $i$.   This delay comprises the queuing delay, and the transmission delay.
 To satisfy the reliability and latency condition, the system needs to keep its average rate larger than average arrival rate, i.e,
\begin{equation}\label{eq:int_rate}
   \lim_{t\to\infty} \frac{1}{T} \int_0^T r_i(t) dt > \lim_{t\to\infty} \frac{1}{T} \int_0^T \lambda_i(t) \beta_i(t) dt,
\end{equation}
where $\beta_i(t)$ is the instantaneous packet size and $\lambda_i(t)$ is the  {instantaneous} packet arrival rate of user $i$ at time-slot $t$. Hence, $\lambda_i(t) \beta_i(t)$ captures the  {instantaneous} arrival rate in bits per second (bps).  (\ref{eq:int_rate}) is essentially a condition on the minimum required rate based on queue stability condition and is not a tight lower bound.
 However, the deep-RL algorithm will use the following tight lower bound:
\begin{equation}\label{eq:implicitRate}
    r_i(t)\geq\phi(\lambda_i(t),\beta(t),\gamma_i,D_i^{\max}),
\end{equation}
 where $\phi(.)$ is an unknown function that we will implicitly approximate  using our proposed method. Our goal is to allocate resources  to minimize the average power while maintaining the users' reliability, latency, and rate of the users, a problem that we formally pose as follows:
\begin{subequations}\label{eq:main_problem}
		\begin{align}
		\min_{p_{ij},\rho_{ij}} \quad &\lim _{t\to \infty} \frac{1}{t} \sum_{\tau=1}^{t} \sum_{i=1}^{N} \sum_{j=1}^{K} p_{ij}(\tau), \label{eq:main_problem:objective_function} \\
		\text{s.t} \quad & \Pro\{D_i>D_i^{\max}\}<1-\gamma_i^*,\quad \forall i \in \mathcal{N},\label{eq:reliabilityCondition}\\
		&  {r_i(t)>\phi(\lambda_i(t),\beta(t),\gamma_i,D_i^{\max}),\quad \forall i \in \mathcal{N},\quad \forall t}\label{eq:RateCondition}\\
		&p_{ij}(t) \geq 0, \quad \rho_{ij}(t) \in \{0,1\},
		\quad\forall i \in \mathcal{N},\quad  \forall j\in\mathcal{K}, \quad \forall t, \label{eq:feasibility}\\
		&\sum_{i}\rho_{ij}(t)=1, \quad \forall j\in\mathcal{K}, \quad \forall t. \label{eq:uniqueness}
		\end{align}
	\end{subequations}
The objective function in (\ref{eq:main_problem:objective_function}) is the average power spent by the BS.
In (\ref{eq:reliabilityCondition}), $D_i$ is the packet delay of user $i$. Constraint (\ref{eq:reliabilityCondition}) takes into account each user's reliability and latency explicitly. the rate constraint is considered both implicitly using (\ref{eq:implicitRate}) and explicitly in (\ref{eq:RateCondition}).  
 Constraint (\ref{eq:reliabilityCondition}) is a reliability condition that guarantees the end-to-end delay to be less than $D_i^{\max}$ with a reliability of at least $1-\gamma_i^*$.   Equation (\ref{eq:reliabilityCondition}) is  the main constraint for ensuring reliable low latency communications. This constraint can capture the reliability and latency, explicitly, as well as the rate and application-specific requirements of user $i$ implicitly. Since providing latency and reliability can vary for different applications and different arrival rates, this constraint is a function of the application type and the arrival rate of packets for the user. Moreover, since there is no universal packet size/arrival model for traffic that can represent countless users/applications altogether, we cannot use a rigid, pre-determined model for handling (\ref{eq:reliabilityCondition}). Therefore, unlike previous works, we do not assume a known user traffic model that can only be applied to a narrow set of applications and, instead, we adopt a model-free approach in which the \ac{drl} agent can adaptively learn each user's traffic model.  Constraints  (\ref{eq:feasibility}) and (\ref{eq:uniqueness}) are feasibility conditions. 
  Note that minimizing power in (\ref{eq:main_problem:objective_function}) does not mean that we sacrifice reliability or latency for minimizing power. In contrast, since we have strict constraints on reliability, latency, and rate, minimizing power cannot sacrifice the QoS, and such an objective is used for the purpose of finding the minimum feasible power for realizing the \ac{urllc} requirements.

 Note that  (\ref{eq:main_problem}) cannot be solved using conventional convex or non-convex optimization problem because: a) without assuming explicit models for the arrival rate and packet size, constraint (\ref{eq:reliabilityCondition}) is not tractable and b) any explicit model for the arrival rates and packet sizes includes  approximations that may not hold true in all practical scenarios. Hence, we adopt an RL-based approach which is more appropriate for the considered setting as we explained later in this section.
  {The key motivation of (3) is to meet specific reliability targets with minimum resource usage, and we chose power as the central resource for our optimization problem. Since reliability is essentially a constraint that a given service seeks to achieve, this constraint inherently captures latency and rate as it is a constraint on the system latency}

At each time slot $t$, the resource allocation system has two functions: \emph{Phase 1}: Determining the rate that each user $i$ should obtain to ensure a target reliability $\gamma^*_i$ and \emph{Phase 2}: Allocating power and RBs to each user so that the power is minimized. Note that the minimum power in Phase 2 is a function of the data rates determined in Phase 1.
To determine the reliability of the system in (\ref{eq:reliabilityCondition}), it is customary to use a specific queuing model, as done in all of the prior art \cite{Bennis2018EVT,Hou2018Bursty, lai2013qos, Zhou2018VTC, bennis2018ultra,yousefvand2018joint,samarakoon2018federated,ashraf2018access,  Kasgari2019TCOM, anand2018infocom}. In contrast, to be model-free, we propose to obtain the reliability in (\ref{eq:reliabilityCondition}) using an empirical measurement of the number of packets transmitted to user $i$ whose delay exceeds $D_i^{\max}$  over the total number of packets transmitted (to user $i$) in time slot $t$, i.e.,
\begin{equation}\label{eq:MeasuredReliability}
\gamma_i(t)=1-\text{Pr}\left\{D_i>D_i^{\max}\right\}\approx 1-\frac{\mu_i'(t)}{\mu_i(t)},
\end{equation}
where $\mu_i'(t)$ is the number of packets transmitted to user $i$ in time slot $t$, whose end-to-end delay exceeds $D_i^{\max}$. $\mu_i(t)$ is the total number of packets transmitted to user $i$ in time slot $t$.   {In order to find delay of each packet, at the source where the packet is generated, a timestamp is added to each packet. After decoding the packet at the destination, its end-to-end latency can be calculated. }
We derive this definition of the reliability based on the 3GPP definition in \cite{3gpp} where reliability is defined as the percentage value of the amount of sent packets that successfully delivered to a user within the time constraint required by the targeted service, divided by the total number of sent packets \cite{Omid2019Magazine}. By doing so, we do not need to make any a priori assumptions on the delay model of the users. Moreover, counting the number of packets is a simple and practical feedback for the network, because each user can convey this information to the BS via a control channel. As $\mu_i(t)$ grows, the approximation converges to the reliability in (\ref{eq:reliabilityCondition}). 
 {We note that \ac{drl} algorithms build a long-term reward function using their \acp{dnn}. Using this long term reward function, a \ac{drl} algorithm can find the long term consequences of its actions. Hence, by using a virtual environment beforehand, one can give the \ac{drl} agent, enough time to find the long term consequences of its actions and train its \ac{dnn} accordingly. Using this virtual environment, the \ac{drl} agent will no longer need significant overhead to achieve high reliability which makes it suitable for our considered \ac{urllc} applications.}
As will be evident from Section \ref{sec:simulResults}, despite having no model for the traffic, the proposed approach will still be able to ensure the target reliability and delay for each user.  {However, there are some practical considerations that must be taken into account when we deploy this system in a real-world, live wireless system. For instance, there will be some sources of delay that our algorithm cannot control, such as the processing delay at the user equipment (UE) and BS or the hardware delay at the base station. However, since our algorithm uses the end-to-end delay as feedback to its learning operation, it will inherently take into account hardware and processing delay at the time of its deployment in a practical system. Indeed, because we rely on a model-free solution and on a GAN-based virtual environment training, our approach can quickly adapt to any new system and allocate resources so that the minimum feasible delay is achieved given available resources and practical network limitations.}

Table \ref{TablePara} provides a list of our main parameters and notations.

 Since the OFDMA resource allocation problem involves a large state space and we have no prior knowledge of the traffic models, we propose a deep-RL framework \cite{deepRL2015} to allocate resources to the users so that their rate requirement and their stringent reliability constraints are satisfied. 
Beyond being able to operate without any model, the key advantage of deep-RL over classical reinforcement learning (RL) is that it can solve control problems with a large state space \cite{deepRL2015}. Deep-RL uses a deep neural network approximating the action-value function in RL.

Similar to humans, gaining valuable experience on the network model can take a long time for a deep-RL agent because extreme network conditions or events will rarely occur in the system. Given that \ac{urllc} requires reliability to such extreme events, it is imperative the our deep-RL agent be equipped  with enough experience on the system. In particular, if we can reproduce these situations for our deep-RL agent in a virtual environment, then it can gain experience rapidly and provide ultra reliability to the network. We propose to generate such an environment for our deep-RL agents by using a GAN-based refiner, as explained  in Section \ref{sec:GAN}.
\begin{table}[!t]
 	\normalsize
 	\begin{center}{
 		\caption{ List of notations.}
 		\scriptsize
 		\label{TablePara}
 		{
 			\begin{tabular}{|c|c|c|c|}
 				\hline
 				\textbf{Notation} & \textbf{Description} &\textbf{Notation} & \textbf{Description} \\ \hline \hline
 				$B$	&    RB bandwidth      &$\ab_t , \mathcal{A}$  &Action and action space \\ \hline
 				
 				$N$	&     Number of users  &$\stb, \mathcal{S}$  &State and state space \\ \hline
 				
 				$K$	&     Number of available RBs  & $\pi_{\psb}(\ab_t|\stb_t)$   & Policy function    \\ \hline
 				$\mu_i(t)$	&    Total  number of packets transmitted to user $i$ in time slot $t$      &  $\hat A_t$ & Advantage function   \\ \hline 
 				$\mu_i'(t)$	&    \makecell{Number of packets transmitted to user $i$ in \\time slot $t$ whose end-to-end delay exceeds $D_i^{\max}$ }  &$F(.;\theta_R)$ &Refiner output \\ \hline
 				$D_i$	&    Packet delay for user $i$   &$z$ &Refiner input   \\ \hline
 				
 				$\beta_i(t)$	&      {Instantaneous}  packet  size for user $i$ at time slot $t$  &$\theta_D$ &Discriminator set of weights \\ \hline
 				
 				$\lambda_i(t)$	&     {Instantaneous}  packet  arrival  rate  of  user $i$ at time slot $t$  &$\theta_R$ &Refiner set of weights  \\ \hline

 				$h_{ij}(t)$	&   Time-varying channel gain for user $i$ on RB $j$    &$D(.;\theta_D)$ &Discriminator output   \\ \hline

 				$\rho_{ij}(t)$	&    RB allocation indicator  for user $i$ on RB $j$  &$\mathcal{N}$ & Set of users  \\ \hline
 				
 				$p_{ij}(t)$	&    Downlink transmission power  for user $i$ on RB $j$ 
 				&$\mathcal{K}$ & Set of RBs\\ \hline

 				$\gamma_i(t)$	&     Reliability for user $i$  & $r_i$ & User $i$ actual rate   \\ \hline

 				$D_i^{\max}$	&      Target end-to-end latency threshold for user $i$ & $r^d_i$ & User $i$ desired rate    \\ \hline
 				
 				$w_i(t)$	&      Time-varying  weight    &$N_0$   & Power spectral density of noise    \\ \hline

 				
 				
 				
 				
 				
 				
 				
 				

 				
 		\end{tabular}}
 	}
 	\end{center}
 	\vspace{-3em}
 \end{table}

\section{Deep-RL for model free \ac{urllc}}\label{sec:deepRLURLLC}
The proposed deep-RL framework will use two feedback inputs to evaluate its performance and update its DNN in each time slot: The total BS downlink power in each time slot
$
     P(t)=\sum_{i=1}^{N} \sum_{j=1}^{K} p_{ij}(t)
$
and the measured reliability of each user at each time using (\ref{eq:MeasuredReliability}). 
Using those two inputs, the deep-RL framework can determine $\rho_{ij}(t)$ and $p_{ij}(t)$ for all $i$ and $j$. After iteratively assigning $\rho_{ij}(t)$ and $p_{ij}(t)$ and receiving the needed feedback in a few time slots, the system can maintain reliability, latency, and rate for each user. 

\begin{figure}[!t]
    \centering
    \includegraphics[trim={.1cm .1cm .1cm .1cm},clip, width=.30\textwidth]{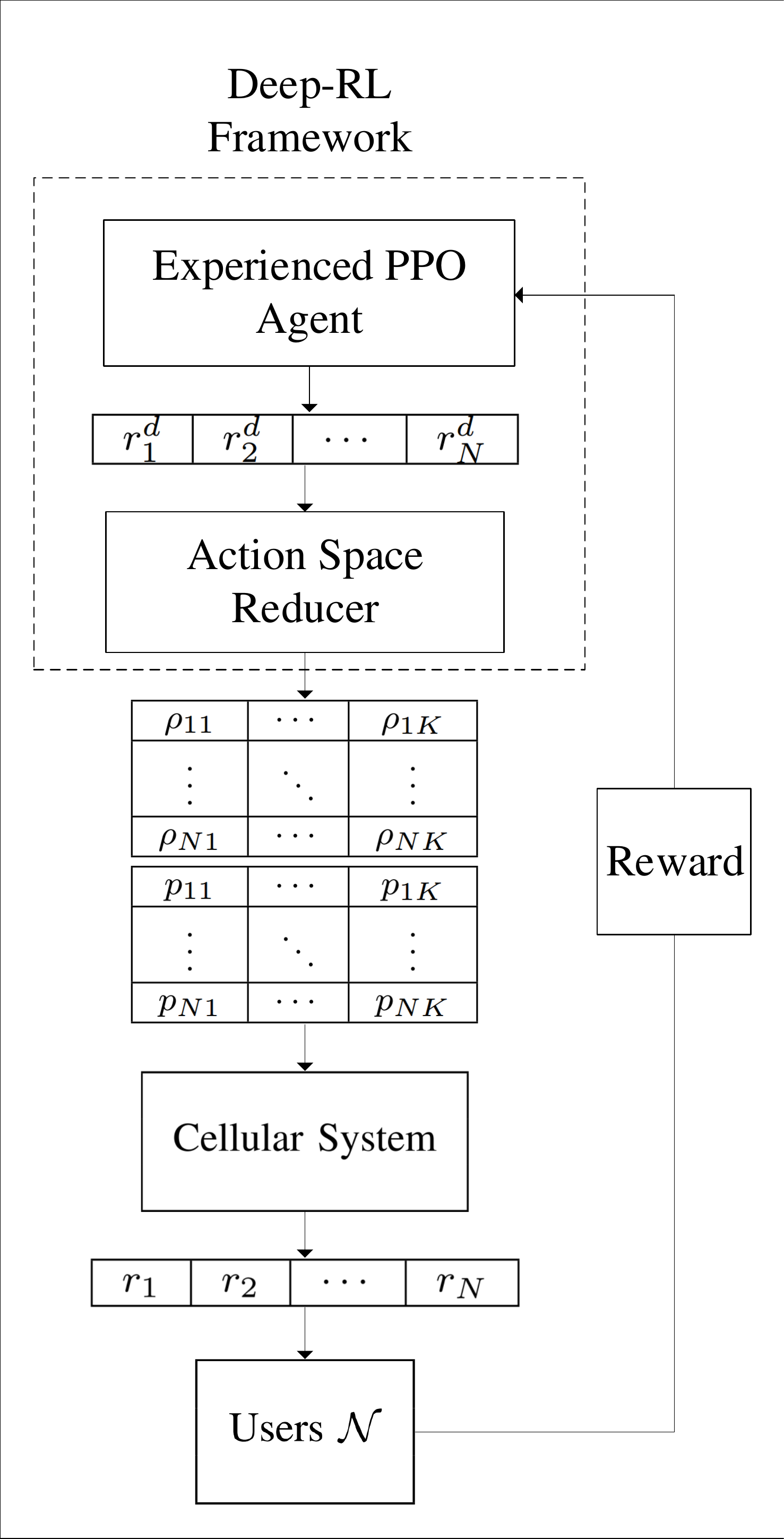}
    \caption{Block diagram for the proposed framework.}
    \label{fig:deepRL}
    \vspace{-2em}
\end{figure}

Our  deep-RL framework is shown in 
Fig. \ref{fig:deepRL}. From Fig \ref{fig:deepRL}, we can see that, at each time slot, the deep-RL algorithm will determine a desired rate $r_i^d(t)$ for each user $i$. Next, an action space reducer maps $r_i^d(t)$ to the OFDMA resources $\rho_{ij}$ and $p_{ij}$ for all $i \in \mathcal{N}$ and $j \in \mathcal{K}$ while minimizing the power (Section \ref{sec:ActionReducer}). Then, each user attains the rate $r_i(t)$ (which is now close to $r_i^d(t)$) and finds a reward function (defined in (\ref{eq:rewardFunc})) and sends it as  feedback to the deep-RL framework that uses this feedback and updates each user's $r_i^d(t)$ accordingly (Section \ref{section:policyG}).
$\vspace{-1.5em}$
\subsection{Deep-RL scheduling}\label{sec:deepRL}
We now formally define our deep-RL framework  by determining  its action-space $\mathcal{A}$, state-space $\mathcal{S}$, and reward function $R$. At each state $\stb_t \in \mathcal{S}$, our deep-RL algorithm takes action $\ab_t \in \mathcal{A}$ and receives a reward $R(\ab_t,\stb_t)$. For our wireless resource allocation problem, we consider the channel gains, the number of packets $\mu_i(t)$ transmitted to each user, and the average packet length $\hat{\beta}_i(t)$ for each user $i$  as the  {\emph{state} $\stb_t=(\hat{\mu_i}(t), \hat{\beta}_i(t), h_{ij}(t)), \forall i \in \mathcal{N}, j \in \mathcal{K}$, for the proposed deep-RL framework. The packets that have user $i$ as their destination arrive with the rate of $\hat{\mu_i}(t)$ at the base station.}  The \emph{action} $\ab_t=(p_{ij}(t), \rho_{ij}(t)), \forall i,j$ is essentially the power and RB allocation for any user $i$ and for any RB $j$.
Then, we determine the reward for deep-RL to guarantee \ac{urllc} without the delay model.
Deep-RL will use this reward for training its DNN and approximating the action-value function.

We define the reward for the proposed deep-RL framework as a function of power and reliability. However, it is implicitly a function of the state and action of the deep-RL algorithm. We define the reward  as follows:
\begin{equation}\label{eq:rewardFunc}
R(\ab_t,\stb_t)=-\sum_{i \in \mathcal{N}} w_i(t) (1-\gamma_i(\ab_t,\stb_t))-\alpha P(\ab_t),
\end{equation}
where $\alpha$  is a weighting factor for power and $P(\ab_t)=\sum_{i=1}^{N} \sum_{j=1}^{K} p_{ij}(t)$.  Here, $\alpha$ can only affect the system before the convergence of \ac{drl} algorithm. This is because, after convergence of the algorithm, $w_i$ will automatically balance the weights between power and the reliability constraint. For notational simplicity, hereinafter, we use $\gamma_i(t)$ and $P(t)$ instead of   $\gamma_i(\ab_t,\stb_t)$  and $P(\ab_t)$. $w_i(t)$ is given by:
\begin{equation}\label{eq:virtualQ}
w_i(t+1)=\max\{w_i(t)+\gamma_i^*-\gamma_i(t),0\}.
\end{equation}
  {$\gamma_i^*$ is the target reliability and $w_i(t)$ is a time-varying weight that increases if $\gamma_i(t)<\gamma_i^*$.}  {Therefore, $w_i(t)$ increases as long as the reliability and accordingly the latency and arrival rate are not guaranteed in the system, i.e, $\gamma_i(t)$ is less than $\gamma_i^*$. Then, the system cannot increase its reward without ensuring that the reliability for each user is guaranteed. Hence, $w_i(t)$ ensures that the system meets the target reliability its users.}  {Since \ac{drl} is used here to find solution to a nonconvex optimization problem using an \ac{sgd} based algorithm, the algorithm is guaranteed to converge to a local minimum \cite{fehrman_convergence_2019}. In the next theorem, we show that, at the convergence point, our algorithm can guarantee reliability.} Hence, we show that, when the proposed \ac{drl} algorithm maximizes the reward  in (\ref{eq:rewardFunc}), the reliability and delay in (\ref{eq:reliabilityCondition}) are guaranteed for each user.
 
\begin{theorem}\label{th:fixd point}
If the BS maximizes the reward in (\ref{eq:rewardFunc}), then after the convergence of the deep-RL algorithm, 
the reliability of each user is guaranteed such that $\gamma_i(t)\geq\gamma^*\,\forall i \in \mathcal{N}$.
\end{theorem}
\begin{IEEEproof}
First,  assume that the value that $w_i(t)$ converges to is $w_i^*$.  Then, we have to show: 
\begin{align*}
&\|w_i(t+1)-w_i^*\|^2=\|\max\{w_i(t)+\gamma_i^*-\gamma_i(t),0\}-w_i^*\|^2=\nonumber \\
& \|w_i(t)+\gamma_i^*-\gamma_i(t)\|^2+\|w_i^*\|^2-2 w_i^{*T}(w_i(t)+\gamma_i^*-\gamma_i(t))=\nonumber\\
&\|w_i(t)-w_i^*\|^2+\|\gamma_i^*-\gamma_i(t)\|^2-2(w_i^*-w_i(t))^T(\gamma_i^*-\gamma_i(t)),
\end{align*}
Hence, 
\begin{align}
&\|w_i(t+1)-w_i^*\|^2-\|w_i(t)-w_i^*\|^2=\nonumber\\
&\|\gamma_i^*-\gamma_i(t)\|^2
-2(w_i^*-w_i(t))^T(\gamma_i^*-\gamma_i(t)).
\end{align}
Hence, if
$
    \|\gamma_i^*-\gamma_i(t)\|^2
<2(w_i^*-w_i(t))^T(\gamma_i^*-\gamma_i(t)), 
$ The algorithm converges to its fixed point.
At the fixed-point of the algorithm, we know that $w_i(t+1)=w_i(t)$, therefore
\begin{equation}
w_i(t)+\gamma_i^*-\gamma_i(t)\leq \max\{w_i(t)+\gamma_i^*-\gamma_i(t),0\}=w_i(t),
\end{equation}
Thus, we can see that
$
w_i(t)+\gamma_i^*-\gamma_i(t)\leq w_i(t)
$
and hence, $\gamma_i(t)\geq \gamma_i^*$.
\end{IEEEproof}
Theorem \ref{th:fixd point} finds the condition for convergence of the algorithm, and ensures that the reliability of each user is guaranteed at the fixed-point of the algorithm, i.e., when $w_i(t)=w_i(t+1)$. Also, the latency for each user is implicitly guaranteed by Theorem \ref{th:fixd point}.
The original action space for the deep-RL resource allocation problem is the possible set for $\rho_{ij}$ and $p_{ij}$ for all $i$ and $j$ which has the size of $\mathcal{O}(K^N)\times \mathds{R}^K$. 
Therefore, in our \ac{urllc} problem, we have a large state space and a large action space. Even though deep-RL effectively addresses the large state space problem, we still have to address the large action space problem. To this end, next, we propose a novel mechanism, called \emph{action space reducer}, using which we reduce the size of the action space. Such an approach to reduce the action space has not been done in prior deep-RL works \cite{ye2018deep} and \cite{He:2017:DRL,Xu2017ICC,Li2018WCNC,ahmed2018deep,azari2018ML,sun2017spawc}. 

\subsection{Reducing Action Space by Optimal Power Allocation}\label{sec:ActionReducer}
The action space for the studied wireless resource allocation problem consists of the following $N\times K$ RB allocation matrix and $N\times K$ power allocation matrix:
\begin{equation*}
    \rhob=
    \begin{bmatrix}
    \rho_{11} &\hdots &\rho_{1K}\\
    \vdots &\ddots &\vdots\\
     \rho_{N1}  &\hdots & \rho_{NK}
    \end{bmatrix}, \quad    \Pb=
    \begin{bmatrix}
    p_{11} &\hdots &p_{1K}\\
    \vdots &\ddots &\vdots\\
     p_{N1}  &\hdots & p_{NK}
    \end{bmatrix}.
\end{equation*}

Our mixed-integer action space size is $\mathcal{O}(K^N)\times \mathds{R}^K$, and it is infeasible to search for the optimal action in such a space. This, in turn, can significantly slow down the convergence of deep-RL algorithms which hinders their applicability for any wireless resource management system.
To address this problem, we propose an action space reducer framework that can reduce the size of action space to $\mathds{R^N}$. This action space reducer essentially, converts the actions taken in $\mathds{R^N}$ to the original mixed-integer action space of $\mathcal{O}(K^N)\times \mathds{R}^K$ which is interpretable in the downlink of an OFDMA system. In Section \ref{section:policyG}, we use the proximal policy optimization (PPO) \cite{PPOmain,PPOcontinous} algorithm to find optimal actions in  $\mathds{R^N}$ space. Our proposed action space reducer can map the actions taken by the PPO algorithm to the original mixed-integer action space using an optimization framework.
Since optimizing the rate allows us to directly control the reliability and latency, we choose the reduced action space to be the rate of each user.  However, each user's set of rates will have many corresponding feasible  power and RB allocation solutions. We choose the allocation solution with minimum power usage. 
Hence, we pose a new optimization problem, called \emph{action space reducer}, whose goal is to map the reduced action space which is the rate for each user to the original action space, i.e., RB and power allocation matrices as output so that the power is minimized. This optimization problem maps the action space of our deep-RL algorithm to the optimization variables in (\ref{eq:main_problem}). To find this RB and power allocation solution, we formally define the action space reducer problem:
\begin{subequations}\label{eq:action_reducer}
\begin{align}
\min_{\Pb,\rhob} \quad &\sum_{i,j} p_{ij}(t)\\
\text{s.t.}\quad &r_i(t)=r_i^d(t),\quad \forall i \in \mathcal{N}, \label{eq:equality_const}\\
&p_{ij}(t) \geq 0, \quad \rho_{ij}(t) \in \{0,1\}, \quad
		\forall i \in \mathcal{N},\quad j \in \mathcal{K} , \quad \forall t,\\
		&\sum_{i}\rho_{ij}=1, \quad \forall j \in \mathcal{K} ,  
\end{align}
\end{subequations}
where constraint (\ref{eq:equality_const}) guarantees that the rate of each user $r_i(t)$ is set to the desired rate for each user $r^d_i(t)$ while minimizing the total BS power.
We can solve (\ref{eq:action_reducer}) with constraint (\ref{eq:equality_const}) in the form of an inequality, i.e., $r_i(t)\geq r^d_i(t)$ using an iterative dual decomposition algorithm. As the number of RBs increases, the primal solution converges to the dual solution and the inequality constraint $r_i(t)\geq r^d_i(t)$ will be satisfied in the form of equality \cite{weiYu2006Tcom, Kasgari2018CISS}. As we will show in  Section \ref{sec:simulResults}, as the number of RBs increases, the error resulting from action space reducer decreases. 

The Lagrangian for problem (\ref{eq:action_reducer}) with inequality constraint $r_i(t)\geq r^d_i(t)$ can be written as:
\begin{equation}
    L(\Pb,\rhob,\lambdab)= \sum_{i,j} p_{ij}(t)-\sum_i \lambda_i(r_i(t)-r^d_i(t)),
\end{equation}
where
   $ \lambdab=\begin{bmatrix}
    \lambda_1 &\lambda_2 &\cdots &\lambda_N
    \end{bmatrix}^T,$
and:
\begin{equation}
	r_i(t)=B \sum_{j=1}^{K} \,\log_2\left(1+\frac{p_{ij}(t) h_{ij}(t)}{\sigma^2}\right).
\end{equation}
The dual problem for (\ref{eq:action_reducer}) is:
\begin{equation}\label{eq:dualproblem}
    \min_{\Pb,\rhob} L(\Pb,\rhob,\lambdab).
\end{equation}
We can see that the dual problem is decomposable for each RB, i.e., we can write (\ref{eq:dualproblem}) as:
\begin{equation}\label{eq:dualSubProblem}
    \min_{p_{ij}(t)}\sum_i p_{ij}(t)-B \sum_i\lambda_i \log_2\left(1+\frac{p_{ij}(t) h_{ij}(t)}{\sigma^2}\right),\quad \forall j\in\mathcal{K}.
\end{equation}
Each subproblem (\ref{eq:dualSubProblem}) is convex and has a closed-form solution. By taking the derivative with respect to $p_{ij}(t)$, we have:
\begin{equation}
    1-\lambda_i B \frac{h_{ij}(t)}{(\sigma^2+p^*_{ij}h_{ij}(t))\log2}=0,\quad \forall i\in\mathcal{N}
\end{equation}

Hence, for each $j$ we have:
\begin{equation}\label{eq:Poewrallocation}
    p^*_{ij}=\left[\frac{\lambda_i B }{\log2}-\frac{\sigma^2}{h_{ij}(t)}\right]^+,\quad \forall i\in\mathcal{N},
\end{equation}
where $[.]^+$ is equivalent to $\max\{.,0\}$. Since each RB can be allocated only to one user, we choose to allocate RB $j$ to user $i_j$ where:
\begin{equation}\label{eq:RBallocation}
    i_j=\argmin_i  p^*_{ij}-B \lambda_i \log_2\left(1+\frac{p^*_{ij} h_{ij}(t)}{\sigma^2}\right),\quad \forall j\in\mathcal{K}.
\end{equation}
Therefore, we find the RB allocated to each user using (\ref{eq:RBallocation}) and the per-RB power using (\ref{eq:Poewrallocation}). The only parameter that remains to be determined is $\lambdab$, which can be derived using the ellipsoid method \cite{weiYu2006Tcom}.  {Since the action space reducer uses the dual decomposition approach, its complexity is $O(KN^3)$ \cite{weiYu2006Tcom}, where $K$ is the total number of resource blocks at each time, and $N$ is the number of users. Hence the algorithm has polynomial time complexity. For deep-RL complexity, one should note that the training is separated from the deployment of \acp{dnn}, and for the complexity of our algorithm, we are mainly focused in deployment. Since the number of users can only affect the last and first layer of the neural network, deployment of a \ac{dnn} has a complexity of $O(1)$, and it is neither scaled with the number of users nor number of resource blocks. Hence, the total complexity of the algorithm will be $O(KN^3)$.}
After reducing the size of action space, each action becomes
    $\ab_t=\begin{bmatrix}
    r^d_1(t) &\cdots &r^d_N(t)
    \end{bmatrix} \in \mathds{R}^N.$ The action space then becomes a $N$-dimensional hyper-cube in $\mathds{R^N}$. This space is not still scalable if used with discretization methods (such as the one in \cite{ye2018deep}). However, since the reduced action space is in $\mathds{R}^N$, instead of discretizing, we can find a solution to our deep-RL problem using PPO, as shown next.
    
    $\vspace{-1.5em}$
\subsection{Optimal Rate Allocation with Policy Gradient}\label{section:policyG}
 After reducing the action space to $\mathds{R}^N$, policy gradient algorithms can effectively control such problems. Among the policy gradient algorithms, we choose PPO, because it has been successfully deployed to cope with high-dimensional continuous action spaces in robotic problems \cite{PPOcontinous}. PPO uses a DNN for mapping the state space into an action space and hence is fast in decision making.  
A policy gradient algorithm such as PPO can determine  $r^d_i(t)$. This is due to the fact that the reduced action space is continuous and using that, we can estimate the gradient of the expected reward of the  policy \cite{PPOmain}. A policy is a conditional distribution function $\pi_{\psb}(\ab_t|\stb_t)$ over a set of actions $\mathcal{A}$, parametrized by $\psi$ that maps each state $\stb_t$ of the system to a distribution.

We use a DNN (with weights $\psb$) to model the policy function $\pi_{\psb}(\ab_t|\stb_t)$. Then, we find $\psb$ using the PPO algorithm \cite{PPOmain,PPOcontinous}. Ultimately, we obtain an optimal policy $\mu_{\psb^*}({\stb}_t)$, which at any given state $\stb_t)$, provides us with the optimal distribution of actions, i.e., $r^d_i(t)$ for each user $i\in\mathcal{N}$. This action $\ab
_t$ is mapped to the RB and power allocation matrices using the action space reducer, and hence, this solves our problem.  {After solving the problem, each user can achieve the rate that is required to satisfy its requirements. These requirements encompass a combination of reliability, latency, and arrival rate that the user needs for its service.}

Now, owing to the PPO and action space reducer, the designed framework can allocate RB and power using a low-complexity and scalable solution. 
While the developed deep-RL framework can be readily deployed for \ac{urllc}, it still needs to be trained in an online manner. Although online training can be done effectively, it can also incur unnecessary overhead as well as long recovery time in case of extreme events. For instance, in case of a sudden rise in arrival rate of all users, the deep-RL system needs a transient time to learn the situation online, i.e.,  recover and allocate resources efficiently. However,  in the \ac{urllc} scenarios, this transient time can be crucial to the performance of the system. To overcome this issue, next, we propose the use of GAN to provide our deep-RL agent with experience in a virtual environment so that it is prepared for the extreme cases encountered by a network of \ac{urllc} users.

$\vspace{-2.5em}$
\section{Experienced Deep-RL with GAN}\label{sec:GAN}
We now propose a generative method to create a virtual environment for training our deep-RL agent and making it experienced. {Here, we need to emphasize that our proposed solution is different from other works such as \cite{liu_data-driven_2020} and \cite{yang_generative-adversarial-network-based_2019} whose goal is to use GAN to approximate a dataset. In contrast, our goal is to use a refiner GAN to \emph{augment} a real dataset with synthetic data in a way to generate a more comprehensive dataset that includes realistic extreme events that were previously unobserved in the real, but limited dataset.}
In particular, we introduce a GAN-based refiner to create a virtual environment that faithfully mimicks a real \ac{urllc} environment. This idea is analogous to the work in \cite{apple2017GAN}, in which the authors used GAN to improve the realism of generating synthetic eye gestures. Inspired by \cite{apple2017GAN}, the proposed GAN will use two inputs: a) A limited set of unlabeled real traffic and channel data and b) Synthetic, simulated data generated based on any standard model for the channel and traffic arrival. Using these two inputs, the proposed GAN-based refiner will create a large and realistic dataset that virtually emulates a realistic wireless environment. In other word, the refined, GAN-generated synthetic channel and traffic data will no longer be distinguishable from real data. Hence, supplementing our real data input with model-based data allows our GAN-based refiner to control the generation of the real-like data. For instance, we can control how fast packets will arrive in the system. \emph{Hence, although the generated data will be similar to real data, we can control it similar to synthetic data.} Using this flexibility in generating real-like data, we can create a virtual environment which is similar to a real \ac{urllc} environment. Then, the deep-RL agent can be deployed in this virtual environment to gain experience before being deployed. By doing so, we can reproduce extreme network conditions and use them to better train our deep-RL agent and provide it with experience on a large-scale, diverse set of network conditions. Here, our deep-RL agent will still operate in a model-free fashion.  In particular, our algorithm can also predict future rewards and decides on how to allocate resources such that reliability is guaranteed in the long term, even without direct short-term feedback.

\begin{figure}[!t]
    \centering
    \includegraphics[trim={0.1cm 0.1cm 0.1cm 0.1cm},clip,width=\mysize\textwidth]{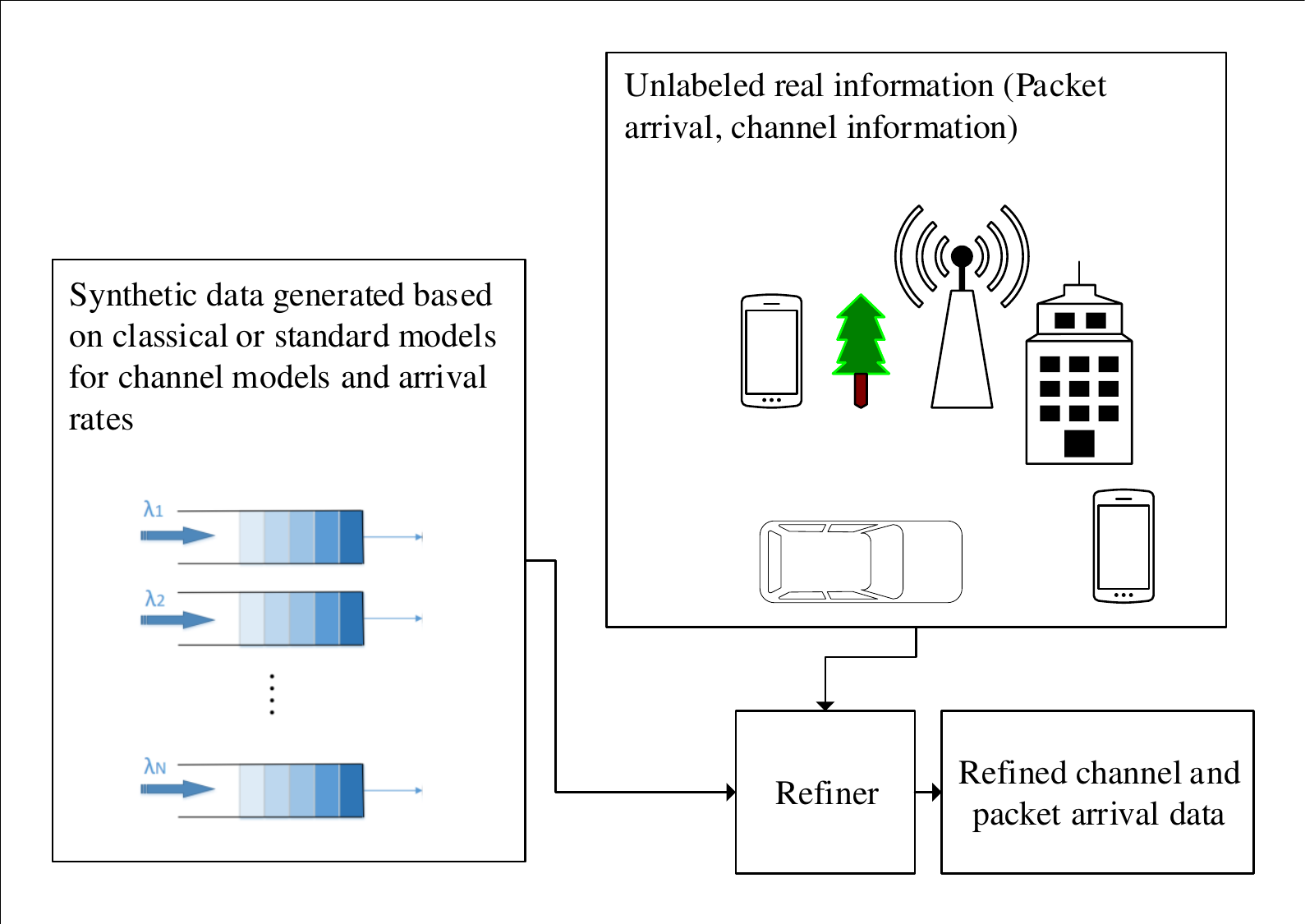}
    \caption{ {Proposed channel and packet information refiner.} }
    \label{fig:Apple}
    \vspace{-2em}
\end{figure}

Fig. \ref{fig:Apple} shows the proposed channel and packet information refiner. The advantage of our proposed refiner over using completely synthetic information is that it removes the bias in the synthetic information thus making the generated channel distribution, packet length distribution, and packet arrival distribution  completely identical to a realistic wireless environment. Moreover, acquiring real datasets on wireless environments is costly, can incur significant overhead on a wireless network, and can raise privacy concerns. Indeed, to date, such real-world datasets remain scarcely available and limited in size. In contrast, using the proposed refiner, without raising any privacy concerns, the network can generate an unlimited set of realistic data at little to no cost and, thus, it will be able to control how many extreme events are in the data. This idea has not been explored in any prior works \cite{He:2017:DRL,Xu2017ICC,Li2018WCNC,ye2018deep,ahmed2018deep},  \cite{azari2018ML,deepRL2015,lillicrap2015continuous,PPOmain,PPOcontinous}, and \cite{apple2017GAN}. Although our work is inspired by \cite{apple2017GAN},  in this work, we control the number of extreme events in the dataset which is not studied in  \cite{apple2017GAN}. Moreover, the approach in \cite{apple2017GAN} is limited to standard image classification tasks and does not leverage GAN for deep-RL enhancement. Finally, we will analytically derive a bound for controlling similarity of the refined data to the input data, a new result that is not done in \cite{apple2017GAN}.

\begin{figure}[!t]
    \centering
    \includegraphics[width=\mysize\textwidth]{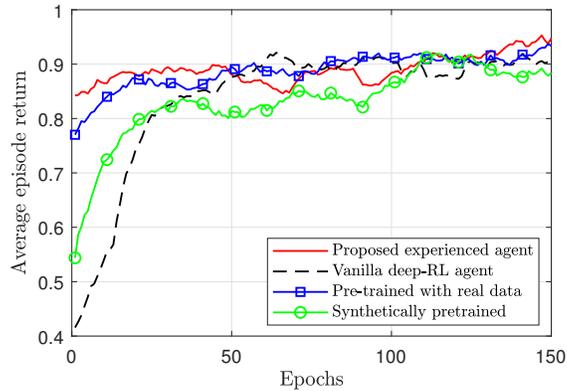}
    \caption{Deploying untrained, experienced, and pre-trained agents in a \ac{urllc} environment.}
    \label{fig:deployingTrainedUntrained}
    \vspace{-2em}
\end{figure}

As we previously mentioned, although our deep-RL agent does not need training and can learn while being deployed, using this training process, the agents become suitable for use in \ac{urllc}. However, if this environment does not resemble the real \ac{urllc} environment, then the deep-RL agent cannot perform well in a realistic environment. This is due to the fact that deep-RL agent needs a transient period to adapt to any new environment. \emph{ Hence, we will use the proposed GAN-based refiner to train our deep-RL agent and turn it into an ``experienced deep-RL agent" before deployment.} To show how the proposed GAN-assisted, experienced deep-RL agent performs, we compare a vanilla deep-RL agent with three pre-trained agents including our experienced deep-RL agent in a wireless environment (having 20 users) in Fig. \ref{fig:deployingTrainedUntrained}. This figure shows how deploying experienced agents that were trained in a virtual environment, a synthetic enviornment, and only real data compare to another agent which  was not pre-trained. As we can see, the  agents trained in real and virtual environments are able to learn faster and have a shorter transition period.  Meanwhile, the untrained agent and synthetically trained agent have longer transient periods during which they do not act optimally. Since \ac{urllc} applications are sensitive to reliability and delay, a short period of unreliability or high delay can have a severe effect on a \ac{urllc} user. Indeed, this non-optimal transient period makes the use of an untrained agent (e.g., as done in existing papers on RL for wireless \cite{He:2017:DRL,Xu2017ICC,Li2018WCNC,ye2018deep,ahmed2018deep}) ill-suited for adoption  in mission critical \ac{urllc} applications. In Section \ref{sec:simulResults}, we will see that training the agent only with real data will remove the transient period, but it will remain susceptible to extreme events. Hence, by designing a GAN-based virtual environment, our goal is to eliminate this transient period when the deep-RL agent is deployed in an actual wireless network, and prepare the agent for extreme events. To the best of our knowledge, no prior work investigated the effect of these transient periods when using deep-RL for wireless networking. 

Designing a virtual wireless environment faces some challenges. If this virtual environment is entirely synthetic, then the transient training period will not be removed. This is due to the fact that even slight changes compared to a  {actual environment} cause the deep-RL agent not to make decision optimally. On the other hand, if this virtual environment is entirely based on collected real wireless data, it does not include enough extreme events. Therefore, this gathered real data does not cover all the unexpected situations for the deep-RL agent. This requires the deep-RL agent to go through a training period when it encounters an extreme event in the \ac{urllc} system. As we observe from Fig. \ref{fig:deployingTrainedUntrained}, this training period from epoch 0 to 50 can cause unreliability and high latency in the \ac{urllc} network. Furthermore, training the agent with synthetic data does not help the agent, and gathering real data for a real \ac{urllc} environment is very time consuming and expensive. 
Hence our designed refiner has two goals: 1) The output of the refiner (packet length, interarrival times, channel gains) must be indistinguishable from a real dataset, i.e., it is from the same distribution as the real dataset and 2) The output of the refiner must have some form of similarity to the input of the refiner. This makes the refiner preserve the main characteristics of the synthetic dataset, e.g., short interarrival time or large packets.

To make the output of the refiner indistinguishable from a real dataset, we use a discriminator neural network along with our refiner neural network \cite{goodfellow2014GAN}.
We assume that the refiner neural network is trained as follows:
\begin{subequations}\label{eq:MainRefinerOptim}
\begin{align}
    \theta^*_R =\quad &\argmin_{\theta_R} \max_{\theta_D} f (\theta_R,\theta_D)=\argmin_{\theta_R} f(\theta_R,\theta^*_D(\theta_R)),\label{eq:RefinerObj}\\
    \text{s.t.} \quad & \Ex_{z\sim \psim}[\|F(z;\theta_R)-z\|]<\epsilon_r, \label{eq:RefinerConstraint}
\end{align}
\end{subequations}
where $\theta_R$ is the set of weights for the refiner neural network, which is the network that gets synthetic data $z$ as input and generates refined real-like data $F(z;\theta_R)$ as output. In our problem, $z$ is equivalent to synthetic state variables $\stb_t$ that we generate for training our deep-RL algorithm. $\theta_D$ is the set of weights of the discriminator, the network that discriminates between the refined data $F(z;\theta_R)$ and real data. $\theta_D^*$ is the optimal set of weights for the discriminator, i.e, the trained discriminator network that can find out if the data is real or not.
$\psim$ is the distribution of synthetic data, e.g., Rayleigh distribution or Poisson packet length distribution. $z$ is a sample from the distribution $\psim$ and $F(z;\theta_R)$ is output of the refiner network. By optimizing the objective in (\ref{eq:RefinerObj}), we make sure that the distribution of the real data and generated data are the same. Moreover,  (\ref{eq:RefinerConstraint}) guarantees the similarity of input of the refiner $z$ and the output of the refiner $F(z;\theta_R)$. If we increase $\epsilon_r$ in (\ref{eq:RefinerConstraint}), we can allow the output of the refiner to differ more from the synthetic dataset and be indistinguishable from the real dataset. On the other hand, if we decrease  $\epsilon_r$, we limit the output of the refiner network to be similar to its input which is the synthetic data. However, if $\epsilon_r$ is less than a certain value, then the optimization problem (\ref{eq:MainRefinerOptim}) becomes infeasible because the refined samples which are noticeably similar to the synthetic dataset are easily identifiable by the discriminator. Next, in Theorem \ref{th:trivial solution}, we rigorously set a insightful guideline for choosing $\epsilon_r$ so that the problem (\ref{eq:MainRefinerOptim}) becomes feasible and also becomes responsive to the input of the refiner (synthetic data). 


\begin{theorem}\label{th:trivial solution}
The optimization problem (\ref{eq:MainRefinerOptim}) is not feasible if  $\epsilon_r<\epsilon_r^t$, where:
\begin{equation}
    \epsilon_r^t=\sqrt{\|\mu_R\|^2+\|\mu_z\|^2-2\mu_R^T \mu_z}.
\end{equation}
$\mu_R$ and $\mu_z$ are the expectations of $F(z;\theta_R^*)$ and $z$, respectively, i.e., $\mu_R=\Ex_{z\sim \psim}[F(z;\theta_R^*)]$ and $\mu_z=\Ex_{z\sim \psim}[z]$. 
\end{theorem}
\begin{IEEEproof}
	See Appendix A.
\end{IEEEproof}


Theorem \ref{th:trivial solution} provides guidelines for choosing the value of $\epsilon_r$.  $\epsilon_r$ must be within a certain range so that the output of the refiner is meaningful. Since refiner is a neural network, it is a flexible function. Hence, if we keep the value of $\epsilon_r$ close to the $\epsilon_r^t$ it will still generate outputs which are entirely similar to the input, and do not have the properties of the real data. Similarly, we can see that by significantly increasing the value of $\epsilon_r$, we will have no control over the output of the refiner, and the output becomes thoroughly similar to the real data.
The constrained refiner optimization problem can be transformed using Lagrange multiplier to:
\begin{align}
    &\theta^*_R =\argmin_{\theta_R} \max_{\theta_D} f (\theta_R,\theta_D)=\argmin_{\theta_R} f(\theta_R,\theta^*_D(\theta_R))+\lambda_r \Ex_{z\sim \psim}[\|F(z;\theta_R)-z\|]\label{eq:GANproblem}\\
    &\theta^*_D(\theta_R) = \text{arg}\max_{\theta_D} f(\theta_R,\theta_D),
\end{align}

where $f$ is a cross entropy loss defined to be:
\begin{equation}\label{eq:GanObj}
    f(\theta_R,\theta_D)= \Ex_{x\sim p_{\text{env}}}[\log(D(x;\theta_D))]+\Ex_{z\sim {\psim}}[\log(1-D(F(z;\theta_R);\theta_D))],
\end{equation}
where $p_{\text{env}}$ is the distribution of real dataset (real wireless environemnt) and $D(x;\theta_D)$ is the output of discriminator neural network parametrized by $\theta_D$ when the input $x$ is given. The term $f(\theta_R,\theta^*_D(\theta_R))$ in (\ref{eq:GANproblem}) is the objective function of the generative refiner which is minimizing the average correct predictions by the discriminator network $D$. This objective function defined in (\ref{eq:GanObj}) is high when the discriminator $D$ is able to discrimante between the real data distributed according to $p_{\text{env}}$ and refined simulated data distributed according to $\psim$. Also, the term $\Ex_{z\sim {\psim}}[\log(1-D(F(z;\theta_R);\theta_D))]$ in (\ref{eq:GANproblem}) guarantees that the output of the refiner is similar to the input of the refiner. $\lambda_r$ in (\ref{eq:GANproblem}) balances the similarity between the input and the output of the refiner and simliarity to the real dataset.

\begin{figure}[!t]
    \begin{minipage}{0.5\textwidth}
		\centering
    \includegraphics[width=\textwidth]{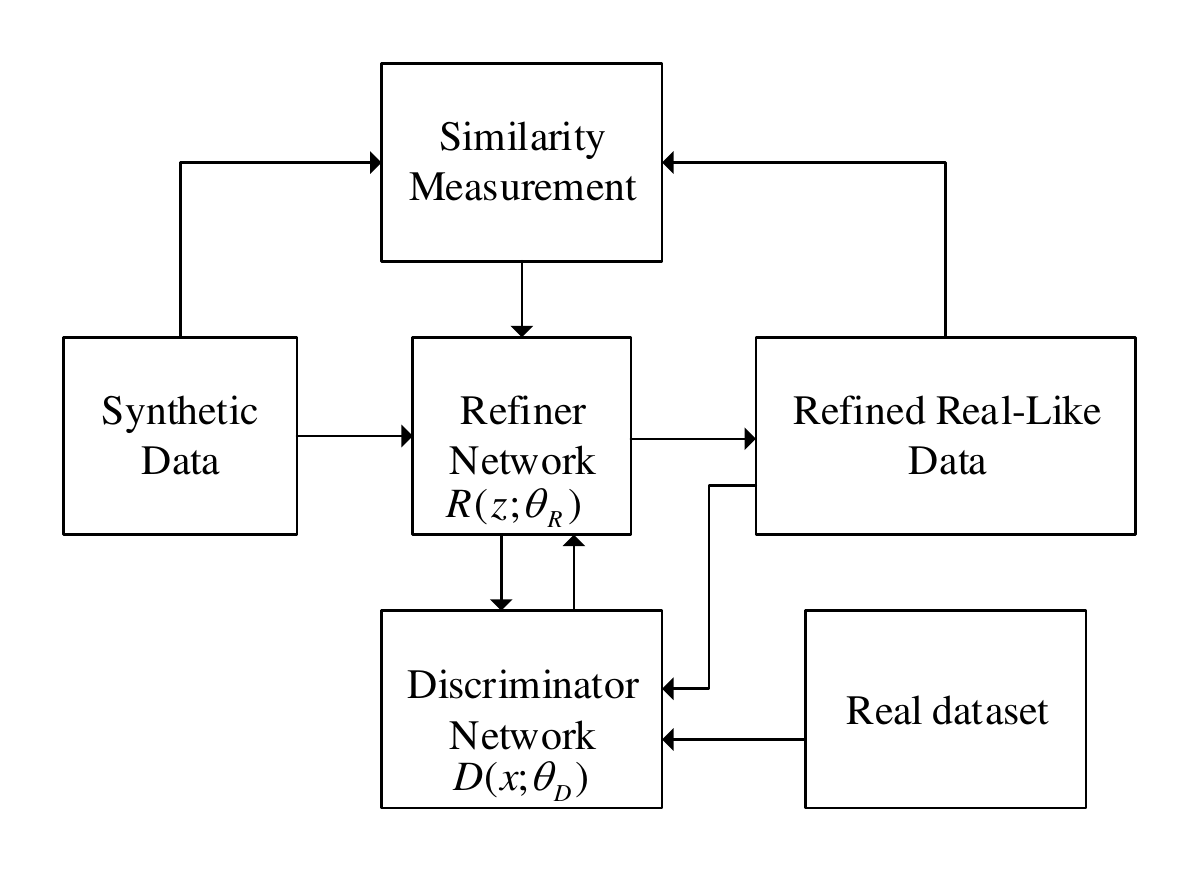}
    \caption{Proposed model for GAN based refiner.}
    \label{fig:RefinerGan}
	\end{minipage}
	\begin{minipage}{0.5\textwidth}s
    	\centering
    \includegraphics[width=\textwidth, trim={0cm 0cm 0cm 1cm}]{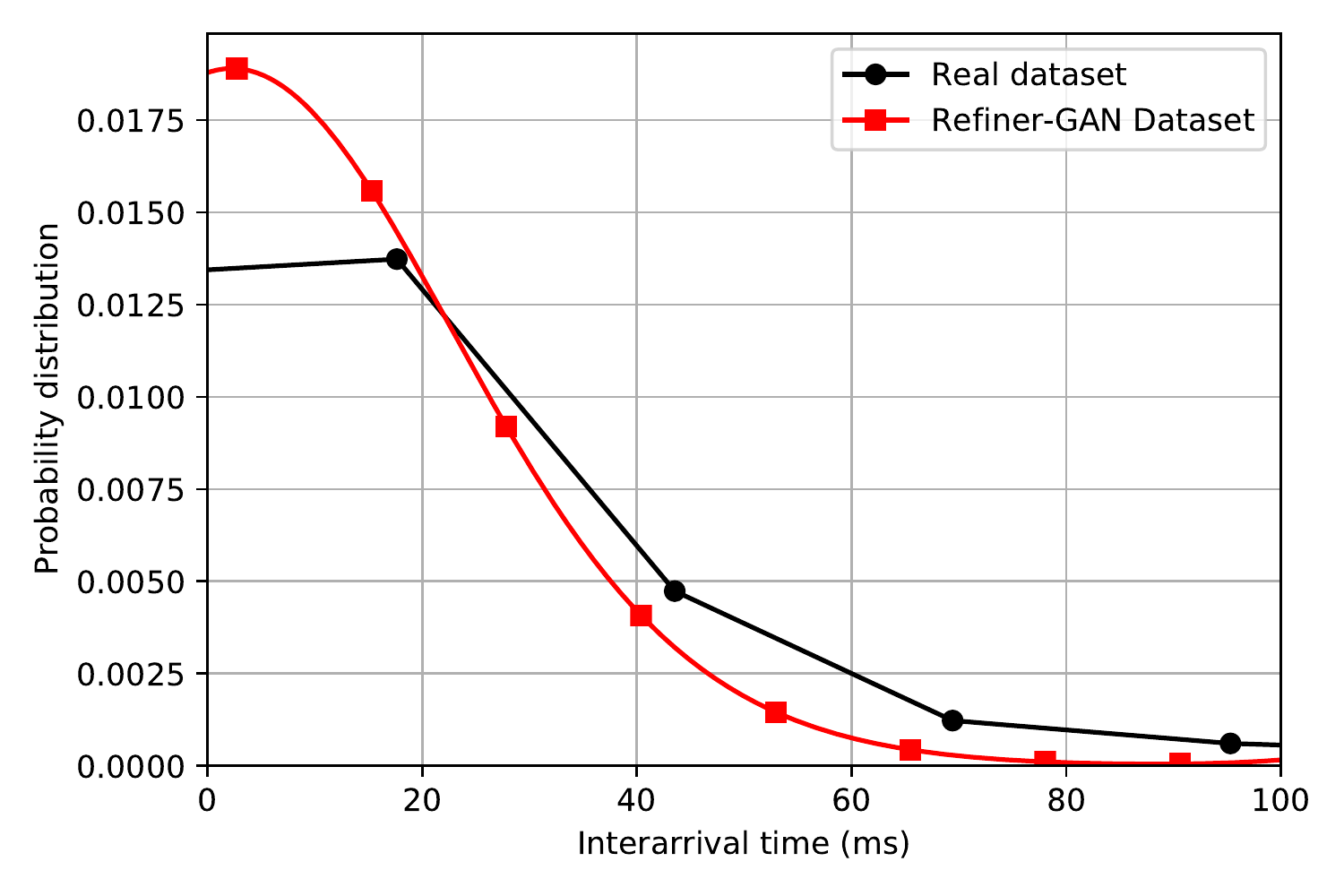}
    \caption{{Probability distribution of generated packet IAT and real packet IAT in the dataset \cite{RojasGC18}.}}
    \label{fig:DistCompare}
    \vspace{-2em}
        \end{minipage}
        \vspace{-2em}
	\end{figure}  
We can see from Fig. \ref{fig:RefinerGan} that the refiner tries to reach its two mentioned goals while the discrimnator tries to discriminates the real data from the refined synthetic data.  The similarity between the input and output of the refiner helps to control the level of extreme events in the generated dataset.
 { Another key challenge is that \ac{drl} can face is the changing number of users in a wireless network.  One possible solution to scenarios with varying network sizes is to train the \ac{drl} agent in an environment in which the number of users is larger than the maximum number of users foreseen in the actual environment. For example, in practical systems, most base stations have a maximum number of users that they can handle. Then, this known maximum capacity can be used to define the neural network of the \ac{drl} agent. In this case, when the network is smaller than the maximum, some users' requirements would be set to zero. These users are then ignored by the system, which means they left the network. Using this workaround, the \ac{drl} algorithm will now allocate resources only to those users having non-zero requirements.}

\section{Simulation Results and Analysis}\label{sec:simulResults}
We simulate a square area of size $500\,\text{m}\times500\,\text{m}$ in which 20 users are served by an OFDMA system with a total bandwidth of $45 \,\text{MHz}$, an RB bandwidth of $B=180$~kHz, and a noise power spectral density $N_0=-173.9\,\frac{\text{dBm}}{\text{Hz}}$, (unless stated otherwise). We set the path loss exponent to 3 (urban area) and the carrier frequency to $2 \,\text{GHz}$.  We set the maximum BS power to $4\,\text{W}$ and each user's latency $D_i^{\max}$ to 10~ms, unless stated otherwise.
For the packet arrival and packet sizes, we used the dataset of \cite{RojasGC18}. Since each row in the  dataset was for a session, we could not access each individual packet size, and interarrival time (IAT). Hence, for each session, we generated the total number of packets with the mean size and IAT mean mentioned
in the dataset in \cite{RojasGC18}.
For evaluation, we assume that the packets arriving from the dataset are similar in length and 
IAT for all users.
However, the proposed deep-RL framework will be model-free and does not have any information about this traffic model.

{Before discussing our key results, in Fig. \ref{fig:DistCompare}, we  show how the proposed refiner-GAN can generate a meaningful dataset out of a limited real dataset. From Fig. \ref{fig:DistCompare}, we can see that the GAN-generated interarrival time (IAT) and the real dataset have a very similar distribution but they are not an exact match. In particular, the packets in the generated dataset tend to have a shorter IAT. This is due to the fact that the refiner-GAN seeks to create a more extreme comprehensive dataset that can help the deep RL train with previously unobserved, extreme points, based on the real dataset.}
$\vspace{-1em}$
\subsection{Experienced Deep-RL Results}
\begin{figure}[!t]
    \centering
    \includegraphics[width=\mysize\textwidth]{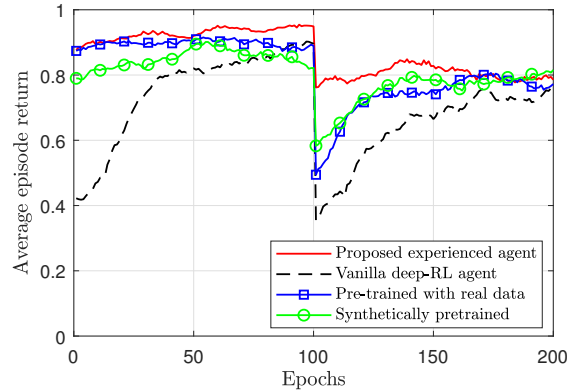}
    \caption{ {Effect of pre-training  the deep-RL agent in multiple enviornments.} }
    \label{fig:ExtremeEvent}
    \vspace{-1.8em}
\end{figure}

Fig. \ref{fig:ExtremeEvent} shows the effect of training an experienced deep-RL agent in the proposed GAN-generated virtual environment, for a network with $10$ users.  {In this experiment, four agents are deployed in a wireless environment in which packets and IAT are generated based on the dataset in \cite{RojasGC18}. We created a virtual environment for training. This virtual environment is a refined version of an M/M/1 arrival with an average packet length of 350 bytes and IAT of 200 microseconds. We considered these values for the packet lengths and IAT as an extreme condition because this IAT value is smaller than 20\% of all packets' IAT in the real dataset, and this packet length value is larger than 78\% of all other packet lengths in the real dataset. We also include the real (not refined) data in the virtual environment. Our experienced deep-RL agent is trained beforehand in the described virtual environment. We compare the experienced agent with three other types of pre-training (as also shown in Fig. 5):
\begin{itemize}
    \item \textbf{Synthetically pre-trained:} Here, the deep-RL agent was pre-trained with completely synthetic (M/M/1) data. This dataset includes extreme synthetic data as mentioned above.
    \item \textbf{Pre-trained with real data:} The deep-RL agent was trained only with the real data.
    \item \textbf{Vanilla deep-RL agent:} The deep-RL agent was not pre-trained.
\end{itemize}
 During the deployment of these four agents, at epoch 100, we change the environment to an extreme environment. This extreme environment has a large packet length and smaller IAT. As we can see from Fig. \ref{fig:ExtremeEvent}, the trained agent can recover around 60 epochs faster than any other agent. This shows that the virtual environment created using the proposed GAN-refiner can make the trained agent ready for real extreme events, and it makes the system resilient to such unpredictable extreme events.} This kind of reliability is an inherent part of \ac{urllc}, and it is especially important for mission-critical applications. A short period of unreliability or long delays can cause irreversible damages to such applications.

\begin{figure}[!t]
    \centering
    \includegraphics[width=\mysize\textwidth]{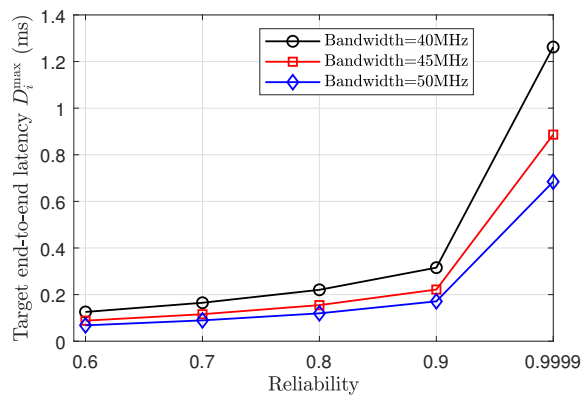}
    \caption{Effect of bandwidth on delay-reliability tradeoff for the proposed system.}
    \label{fig:DmaxRelBandwidth}
    \vspace{-2em}
\end{figure}
Fig. \ref{fig:DmaxRelBandwidth} shows the effect of the maximum bandwidth on the delay-reliability tradeoff. Here, we use the target latency $D_i^{\max}$ because it is the treshold for our reliability analysis. Clearly, as we allocate more bandwidth to the system, we can achieve higher reliability and lower latency with the same rate. For instance, by increasing the bandwidth from 45~MHz to 50~MHz, we are able to decrease the latency of each user by 16\%. Also, we can see that increasing the target latency increases the reliability in the system. This is because larger target latencies makes it easier for the system to satisfy the associated (looser) reliability requirements. Fig. \ref{fig:DmaxRelBandwidth} shows that our system can achieve a reliability of $99.99\%$ and latency of less than $1.5$~ms.

\begin{figure}[!t]
    \centering
    \includegraphics[width=\mysize\textwidth]{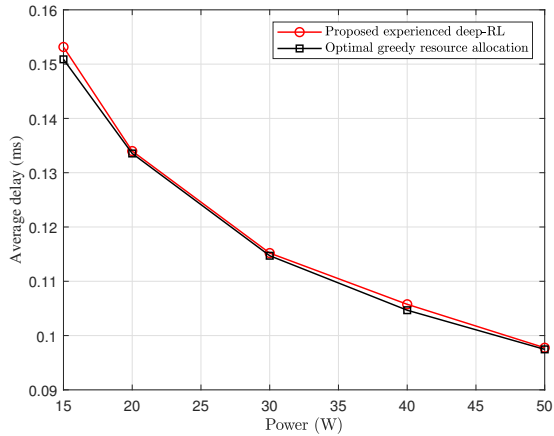}
    \caption{ {Effect of BS power on the average delay of the proposed method and an optimal resource management system.}}
    \label{fig:ActualDelay2}
    \vspace{-2em}
\end{figure}

 {Fig. \ref{fig:ActualDelay2} shows the effect of the BS power on the average delay of the system. For performance assessment, we used a greedy exhaustive search algorithm to benchmark proposed experienced deep-RL agent. In this algorithm, we try all possible resource allocation case in order to find the minimum average delay with the same amount of power. 
 Clearly, our proposed method can reach a near-optimal latency without assuming any model or any specific packet arrival structure.
} Moreover, our system can achieve average delays as low as $0.15$ ms (for small packets) which is suitable for many ultra low latency systems.

\subsection{Deep-RL algorithm}
Next, we provide a set of simulations in which we use only synthetic data in order to showcase the performance of our proposed deep-RL algorithm over a broad range of parameters which are not available within the existing, limited real datasets.

\begin{figure}[!t]
    \centering
    \includegraphics[width=\mysize\textwidth]{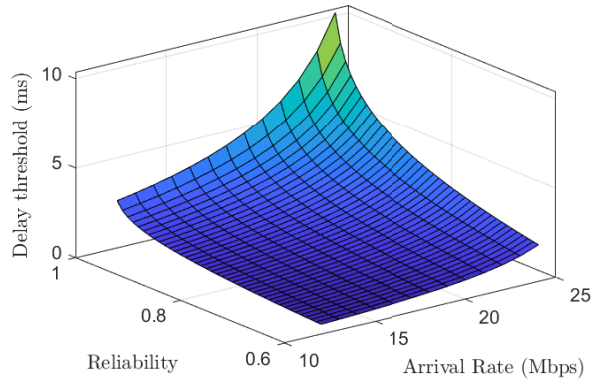}
    \caption{3D plot of the achievable arrival rate, delay, and reliability for the system.}
    \label{fig:3DPlot}
    \vspace{-2.5em}
\end{figure}

Fig. \ref{fig:3DPlot} shows the relation between the arrival rate (i.e., minimum required data rate for a user), maximum delay, and reliability in our system. As we mentioned, the rate, reliability, and latency are incompatible design parameters. However, since our system can attain any feasible combination of  the rate, reliability, and latency, we can enable \ac{urllc} with reliability of 99\% and latency of 4.2~ms with the rate of 1~Mbps, and a reliable high-rate communication with 99\% reliability and rate of 10 Mbps with latency of 24.5~ms at the same time. Also, the system can balance between rate, reliability, and latency. As an example, we can see from Fig. \ref{fig:3DPlot} that, our system is able to provide ultra-reliable low latency communication with a delay of 8 milliseconds and a reliability of $98\%$ with a rate of $7$~Mbps. However, if we need higher rate for this system, without decreasing reliability or increasing latency, then we have to allocate a higher bandwidth to the system. We can increase the rate of each one of the 20 user to $46$~Mbps if we increase the system bandwidth from 45~MHz to 200~MHz and the power from 5~W to 20~W. These results provide insightful guidelines for controlling the rate-reliability-delay tradeoff.
For example, we see that with a reliability of 98\%, delay of 8 ms, and rate of 7~Mbps, a gain of 1\% reliability can be done with 47\% less delay but at the expense of a seven-fold decrease in the rate.

\begin{figure}[!t]
    \begin{minipage}{0.5\textwidth}
		\centering
		\includegraphics[scale=.7,trim={0.6cm 0 0 0}]{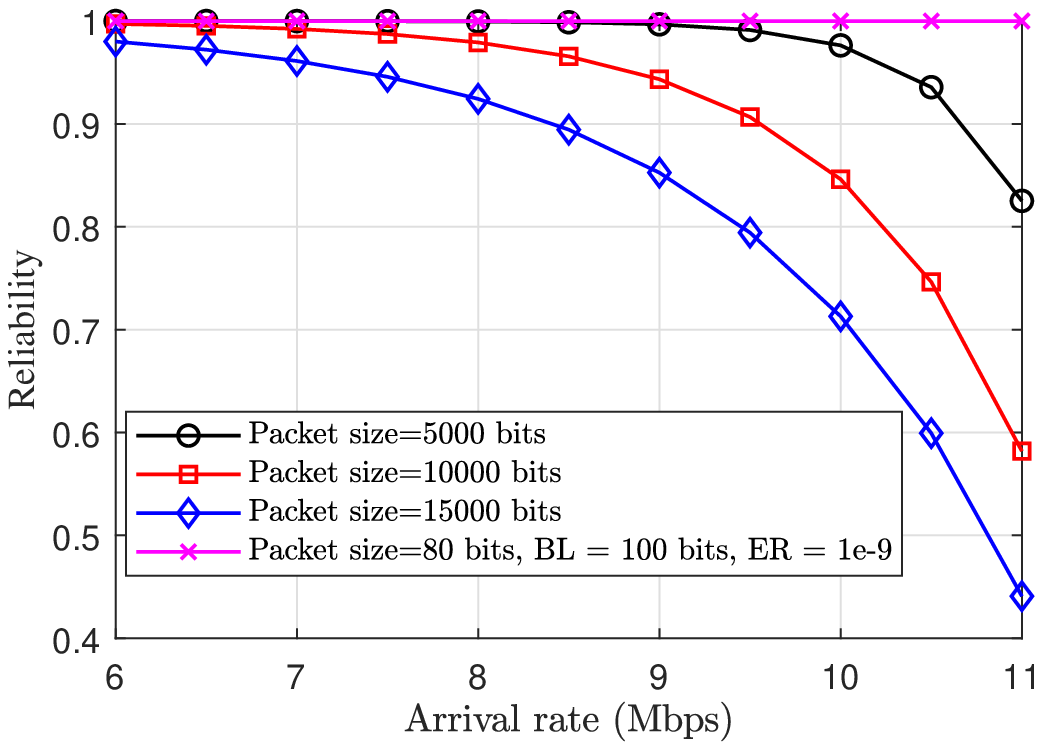}
	\end{minipage}
    	\begin{minipage}{0.5\textwidth}
		\centering
		\includegraphics[scale=.7,trim={0.6cm 0 0 0}]{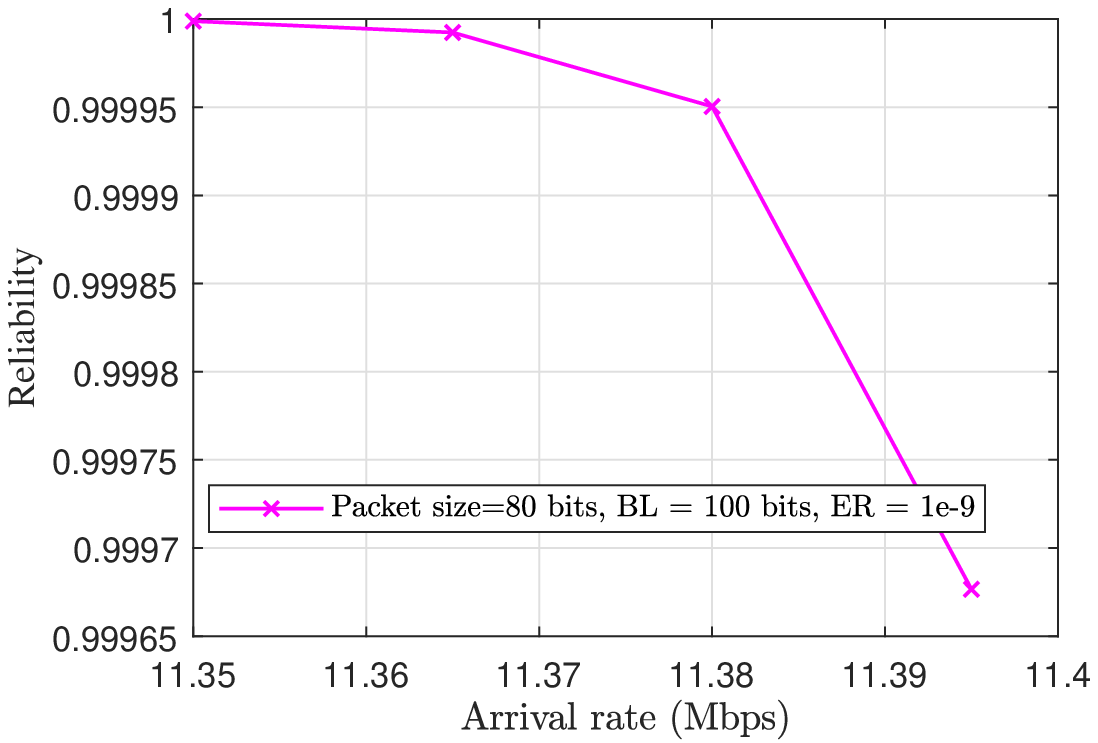}
        \end{minipage}
        \caption{(Left) Effect of packet size on rate-reliability tradeoff for the designed system. (Right) Effect of arrival rate on the reliability within a finite blocklength regime. BL: Blocklength, ER: Error rate}
        \label{fig:DmaxRelPSizer}
        \vspace{-2.5em}
	\end{figure}

The effect of packet size on the reliability of the system with  $D_i^{\max}=$ 10~ms is shown in Fig. \ref{fig:DmaxRelPSizer}. We can see that, for higher rates, the effect of packet size becomes more dominant. The system can only provide more reliability to the traffic with shorter packet sizes. This is due to the fact that applications with shorter packets naturally have a smaller end-to-end delay. Hence, it is less challenging for our system to provide ultra high  reliability to such applications. Fig. \ref{fig:DmaxRelPSizer} shows that our system is able to reach \ac{urllc} reliability and latency as well as higher data rates with moderate latency and reliability for high data rate and large packet size applications. We can see that at higher data rates the reliability decreases, and this is because the limited bandwidth and power in the system can guarantee reliability up to a certain rate. We can increase this reliable rate by either decreasing packet size, increasing bandwidth, increasing power, and/or increasing target end-to-end latency.  {Furthermore, in order to obtain the rate of the users for finite blocklengths regimes, instead of using the rate in (1), we have used the finite blocklength model from \cite{poor_blocklength_2010}. We considered the blocklength to be $100$~bits and the packet size to be $80$~bits and the error rate to be $10^{-9}$. As we can see from Fig. \ref{fig:DmaxRelPSizer}, the reliability of the users having a finite blocklength will be close to 1. This is due to the fact that, although the normal approximation decreases the effective rate of each user, with shorter packets, providing reliability for each user becomes an easier task for the proposed \ac{drl} solution. Clearly, the proposed approach can provide reliability within the \ac{urllc} requirements. In order to further assess the performance of our algorithm in the finite blocklength regime, we have increased the arrival rate to $11.4$~Mbps in Fig. \ref{fig:DmaxRelPSizer} (right) which shows a decrease in the reliability when the arrival rate is closer to the departure rate of the users. From this figure, we can see that the system can still provide a reliability of $99.995\%$ with the arrival rate of $11.38$~Mbps, which is close to the \ac{urllc} requirements. }

\begin{figure}[!t]
    \centering
    \includegraphics[width=\mysize\textwidth]{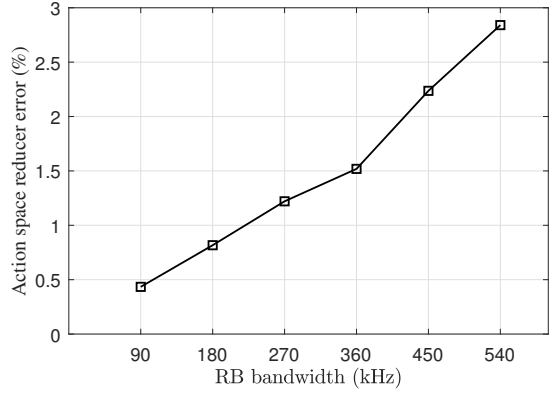}
    \caption{Effect of resource block bandwidth on the per user error of the action reducer.}
    \label{fig:Action_Reducer}\vspace{-2.5em}
\end{figure}

Fig. \ref{fig:Action_Reducer} shows the per user error of the action reducer defined as
$
    E=\frac{\|\rb- \rb^d\|}{N\|\rb\|},
$
where $\rb$ is the vector of wireless downling downlink rate and $\rb^d$ is the vector of desired rate. This error measures the distance between the input and output rate of action space reducer. We can see that, as the bandwidth of each RB decreases, the number of RBs increases in the system, and hence, the error of our action space reducer will decrease. We can see that, for an RB bandwidth of 180~kHz (typical for LTE), the error is less than 1\% for each user.

\section{Conclusion}\label{sec:conclusion}
  In this paper, we have proposed a novel \emph{experienced} deep-RL framework to provide model-free \ac{urllc} in the downlink of an OFDMA system. Our proposed deep-RL framework can guarantee high end-to-end reliability and low end-to-end latency, under explicit data rate constraints without any models of or assumptions on the users' traffic. In particular, to enable the deep-RL framework to account for extreme network conditions and operate in highly reliable systems, we have proposed a new approach based on GANs, namely, GAN-based refiner. We have used this GAN-based refiner to pre-train the deep-RL agent utilizing a mix of real and synthetic data. Using this GAN-based refiner, we can expose our deep-RL agent to a broad range of network conditions.
We have also shown that our proposed approach can predict the users' traffic using the experienced deep-RL framework and subsequently use those predictions in the resource allocation process. We have formulated the problem as a power minimization problem under reliability, latency, and rate constraints. To solve this problem, first, we have determined the rate of each user using experienced deep-RL. Then, using a proposed action space reducer, we have mapped these rates to the resource block and power allocation vectors of a wireless system. Finally, we have used the end-to-end reliability and latency of each user as feedback to the deep-RL framework. We have demonstrated that, at the fixed-point of the deep-RL algorithm, the reliability and latency of the users are guaranteed.
 Moreover, we have derived some analytical bounds for the output of the proposed GAN-based refiner.
 Our results have shown that the proposed approach can achieve the required performance in the rate-reliability-latency region. We have also demonstrated that the proposed experienced deep-RL framework is suitable for \ac{urllc} applications because it can: a) remove the  transient training time that exists in conventional deep-RL methods and b) recover faster in the case of unexpected extreme events which cause failure in \ac{urllc} systems. The proposed experienced deep-RL framework can also be adopted in many future wireless applications that require adaptive and quick optimization algorithms. 

$\vspace{-3em}$

\appendices
	
	\section{Proof of Theorem \ref{th:trivial solution}}
	\begin{IEEEproof}
Assuming the cross entropy loss function, we have:
\begin{align}
     f(\theta_R,\theta_D)&=  \Ex_{x\sim p_{\text{env}}}[\log(D(x;\theta_D))]+\Ex_{z\sim {\psim}}[\log(1-D(F(z;\theta_R);\theta_D))]\nonumber\\
     &=\int_x [\log(D(x;\theta_D))]p_{\text{env}}(x) dx + \int_z [\log(1-D(F(z;\theta_R);\theta_D))] \psim(z) dz\nonumber\\
     &=\int_x [\log(D(x;\theta_D))]p_{\text{env}}(x) + [\log(1-D(x;\theta_D))] p_{r}(x) dx,
\end{align}
where $p_r(x)$ is distribution of the refiner outputs. We know that for all the functions $c$ in the form of $c(v)=a\log(v)+b\log(1-v)$ and any $(a,b)\in \mathds{R}^2\setminus\{0,0\}$, the maximum in the set $v\in[0,1]$ is at $v=\frac{a}{a+b}$. Hence, we know that 
\begin{align}
    \max_{\theta_D} f (\theta_R,\theta_D)&= \int_x [\log(\frac{p_{\text{env}}(x)}{p_{\text{env}(x)}+ p_{r}(x)})]p_{\text{env}}(x) + [\log(1-\frac{p_{\text{env}}(x)}{p_{\text{env}(x)}+ p_{r}(x)})] p_{r}(x) dx\nonumber\\
    &= \text{KL}\left(p_{\text{env}}\|{p_{\text{env}(x)}+ p_{r}(x)}\right)+\text{KL}\left(p_{r}(x)\|p_{\text{env}(x)}+ p_{r}(x)\right), \label{eq:JSD}
\end{align}
    We know that right hand side of (\ref{eq:JSD}) is minimized when $p_{\text{env}(x)}=p_{r}(x)$. However, since (\ref{eq:MainRefinerOptim}) is a constrained optimization problem 
    we should consider conditions other than optimality to find its optimum.  
Considering the complementary slackness condition in the  Karush-Kuhn-Tucker (KKT) conditions for (\ref{eq:MainRefinerOptim}), we know that the final solution, at least one of following two conditions should be satisfied. 1) The final solution of (\ref{eq:RefinerObj}) should satisfy the (\ref{eq:RefinerConstraint}), i.e.,
$
    \Ex_{z\sim \psim}[\|F(z;\theta_R^*)-z\|]=\epsilon_r, 
$
or 2) the optimal solution is same as the solution of  unconstrained  version of (\ref{eq:MainRefinerOptim}) which is
$p_{\text{env}(x)}=p_{r}(x)$.

In case 2) when  $\epsilon_r\to \infty$, the problem becomes unconstrained optimization and the refiner works as if its only goal is to generate more realistic samples instead of controlling the outputs of the refiner. 
However, in case 1) we have:
\begin{align}\label{eq:traceCov}
    \Ex_{z\sim \psim}[\|F(z;\theta_R^*)-z\|^2]
    &=\Ex_{z\sim \psim}[(F(z;\theta_R^*)-z)^T (F(z;\theta_R^*)-z]\nonumber\\
    &=\tr(\Ex_{z\sim \psim}[(F(z;\theta_R^*)-z)^T (F(z;\theta_R^*)-z])\nonumber\\
    &=\Ex_{z\sim \psim}[\tr((F(z;\theta_R^*)-z)^T (F(z;\theta_R^*)-z)]\nonumber\\
    &=\Ex_{z\sim \psim}[\tr((F(z;\theta_R^*)-z) (F(z;\theta_R^*)-z)^T]\nonumber\\
    &=\Ex_{z\sim \psim}[\tr(F(z;\theta_R^*)F(z;\theta_R^*)^T+z z^T -2 F(z;\theta_R^*) z^T)]\nonumber\\
    &=\tr(\Ex_{z\sim \psim}[F(z;\theta_R^*)F(z;\theta_R^*)^T])+\tr(\Ex_{z\sim \psim}[z z^T]) \nonumber\\ &-2 \tr(\Ex_{z\sim \psim}[F(z;\theta_R^*) z^T])\nonumber\\
    &=\tr(\VAR(F(z;\theta_R^*)+\mu_R \mu_R^T)+\tr(\VAR(z)+\mu_z \mu_z^T) \nonumber\\ &-2 \tr(\COV[z,;\theta_R^*) z]+\mu_z\mu_R^T)\nonumber\\
    &=\tr(\VAR(F(z;\theta_R^*))+\tr(\VAR(z))-2 \tr(\COV(z,F(z;\theta_R^*)) \nonumber\\ &+\tr(\mu_R^T\mu_R)+\tr(\mu_z^T\mu_z)-2\tr(\mu_R^T \mu_z )\nonumber\\
    &=\tr(\VAR(F(z;\theta_R^*))+\tr(\VAR(z))-2 \tr(\COV(z,F(z;\theta_R^*)) \nonumber\\ &+\|\mu_R\|^2+\|\mu_z\|^2-2\mu_R^T \mu_z, 
\end{align}
where $\tr(\cdot)$ is the trace of a matrix, $\mu_R$ and $\mu_z$ are the expectations of $F(z;\theta_R^*)$ and $z$, respectively, i.e., $\mu_R=\Ex_{z\sim \psim}[F(z;\theta_R^*)]$ and $\mu_z=\Ex_{z\sim \psim}[z]$. $\COV(.)$ and $\VAR(.)$ are cross-covariance and covariance matrix of random vectors, respectively.The operations in (\ref{eq:traceCov}) are using the fact that both $\Ex_{z\sim \psim}(.)$ and $\tr(\cdot)$ are linear operators and using the properties of function $\tr(\cdot)$.
Using the Cauchy-Schwarz inequality, we know that \begin{equation}\label{eq:Cauchy}
    \tr(\VAR(F(z;\theta_R^*))+\tr(\VAR(z))-2 \tr(\COV(z,F(z;\theta_R^*))\geq 0.
\end{equation}
Inequality (\ref{eq:Cauchy}) is tight when where the random vectors $\VAR(F(z;\theta_R^*)$ and $z$ are completely correlated. Hence, we can see that
\begin{equation}
    \Ex_{z\sim \psim}[\|F(z;\theta_R^*)-z\|^2]=\epsilon_r^2 \geq \|\mu_R\|^2+\|\mu_z\|^2-2\mu_R^T \mu_z,
\end{equation}
and hence the minimum value for the $\epsilon_r$ is as follows:
\begin{equation}
\epsilon_r \geq \sqrt{\|\mu_R\|^2+\|\mu_z\|^2-2\mu_R^T \mu_z,}=\epsilon_r^t.
\end{equation}


\end{IEEEproof}
$\vspace{-3em}$	

	\nocite{*}
	\bibliographystyle{IEEEtran}
	\def\baselinestretch{1}
	\bibliography{PaperBib.bib}

\end{document}